\documentclass[11pt,letterpaper]{article}
\usepackage{axodraw2}
\pdfoutput=1
\usepackage{amscd}
\usepackage{amsmath, bbm} 

\newcommand{\nn}{\nonumber}
\usepackage{jheppub} 

\title{An Anomaly-free Atlas: charting the space of flavour-dependent
  gauged $U(1)$ extensions of the Standard Model} 

\author[a]{B C Allanach**,}
\author[a]{Joe Davighi,\footnote{Corresponding author.}}
\author[a,b]{and Scott Melville}
\affiliation[a]{DAMTP, University of Cambridge, Wilberforce Road, Cambridge, 
CB3 0WA, United Kingdom}
\affiliation[b]{Emmanuel College, University of Cambridge, St Andrew's Street, Cambridge, CB2 3AP, United Kingdom}

\emailAdd{B.C.Allanach@damtp.cam.ac.uk}
\emailAdd{jed60@cam.ac.uk}
\emailAdd{scott.melville@damtp.cam.ac.uk}
\preprint{DAMTP-2018-41}
\abstract{Spontaneously broken, flavour-dependent, gauged $U(1)$ extensions of
  the Standard Model (SM) have many phenomenological uses. We chart the space of
  solutions to the gauge anomaly cancellation equations in such
  extensions, for both the SM chiral fermion content and the SM plus (up to) three right-handed neutrinos (SM$\nu_R$). 
  Methods from Diophantine analysis allow us
  to efficiently index the solutions arithmetically, and produce the complete solution space in particular cases.
  In order to solve the general case, we build a computer program which cycles through possible  $U(1)$ charge assignments, 
  providing  all solutions for charges up to some pre-defined maximum absolute 
  charge. Lists of anomaly-free $U(1)$ charge assignments result, which corroborate the results of our Diophantine analysis.
We make these lists, which may be queried for further desirable properties, publicly available. This
previously uncharted space of anomaly-free charge assignments has been little explored until now, paving the way for future model building
and phenomenological studies.}

\begin{document} 
\maketitle
\flushbottom

\section{Introduction}

Spontaneously broken, gauged $U(1)$ extensions of the Standard Model (SM) are
 currently enjoying a high level of interest in
particle physics, thanks to their ability to answer various phenomenological
questions. For example, they have
been successfully employed to model dark  
matter~\cite{Okada:2010wd,Allanach:2015gkd,Okada:2016tci,Okada:2016gsh,Okada:2017dqs,Agrawal:2018vin,Okada:2018tgy}, to explain measurements of
the anomalous magnetic moment of the muon~\cite{Heeck:2011wj}, to provide
axions~\cite{Berenstein:2010ta} or leptogenesis~\cite{Chen:2011sb}, to explain
the stability of the proton in supersymmetric models~\cite{Carone:1996nd}, to
break supersymmetry~\cite{Kaplan:1999iq},
and to provide fermion masses
through the Froggatt-Nielsen 
mechanism~\cite{Froggatt:1978nt}, to name but a few. 

\paragraph{Flavour non-universality:}
In many of these examples, fermions are given
family-dependent 
$U(1)$ charges. 
A notable recent impetus comes from LHCb measurements 
of lepton flavour non-universality in certain rare neutral current $B$-meson
decays~\cite{Aaij:2014ora,Aaij:2017vbb,Hiller:2003js}.  
{\em Prima facie}, there
are two classes of new particle 
which might be responsible for such an effect at tree-level: a leptoquark, or a new
charge-neutral heavy vector boson (called a $Z^\prime$).
In $Z^\prime$ models for the $B$-meson
decays~\cite{Gauld:2013qba,Buras:2013dea,Buras:2013qja,Altmannshofer:2014cfa,Buras:2014yna,Crivellin:2015mga,Crivellin:2015lwa,Sierra:2015fma,Crivellin:2015era,Celis:2015ara,Greljo:2015mma,Altmannshofer:2015mqa,Allanach:2015gkd,Falkowski:2015zwa,Chiang:2016qov,Becirevic:2016zri,Boucenna:2016wpr,Boucenna:2016qad,Ko:2017lzd,Alonso:2017bff,Alonso:2017uky,1674-1137-42-3-033104,CHEN2018420,Faisel:2017glo,PhysRevD.97.115003,Bian:2017xzg,PhysRevD.97.075035,Bhatia:2017tgo,Allanach:2017bta,Allanach:2018odd,Duan:2018akc},
the $Z^\prime$ arises as the new 
heavy gauge boson from a spontaneously broken $U(1)$
extension to the SM gauge symmetry, under which the charges of chiral fermions are
family-dependent. 

Rather than focus on a particular Beyond the Standard Model scenario, or a
 particular realisation of breaking an additional $U(1)$ group, we shall
 consider the SM as a low-energy Effective Field Theory (EFT) in which the
 fermions may have (in addition to their usual quantum numbers) a
 family-dependent charge under this $U(1)$ gauge group. This approach allows
 us to remain agnostic about the heavy gauge boson which mediates the
 interaction and therefore captures the relevant low-energy
 phenomenology of a wide class of different models.     

\paragraph{Anomaly cancellation:}
If such EFTs are to be embedded into a renormalisable, ultra-violet (UV) completion, then
the additional gauge symmetry (which we shall call $U(1)_F$ from now on)
should be non-anomalous. This means that the
$U(1)_F$ charges of the chiral 
fermions in the theory must be chosen such that all of the anomaly coefficients
cancel, including for the mixed anomalies involving $U(1)_F$, and the
gauge-gravity anomaly. 
The solutions to these highly non-trivial constraints on the 
possible $U(1)_F$ charges of the SM fermions are the subject of this
paper. {\em Our central aim is to categorise and list the 
sets of fermionic charges that solve
the anomaly constraints}. By doing so, we hope to provide inspiration for
model building and aid future phenomenological studies. In addition to the SM fermions,
we shall also include the possibility of three heavy right-handed (RH) neutrinos, since it is a
popular minimal extension that can explain the size
of neutrino masses inferred from
neutrino oscillation data. 
The ``anomaly-free atlas'' of $U(1)_F$ charges is stored on {\tt
  Zenodo} at \url{http://doi.org/10.5281/zenodo.3345889}~\cite{zenodo}.

\paragraph{Wess-Zumino terms:}
Before we elaborate on the form these constraints take, and sketch
how we solve them, we would like to comment on the role of anomaly
cancellation in realistic model building, in which low-energy theories are
necessarily regarded as ``only'' EFTs, and are {\em not}\/
intended as fundamental theories. In this case, it is of course feasible that
anomalies do not cancel in the low-energy EFT, but are cancelled at high energies
by new UV physics. 
 For example, heavy chiral fermions may have been integrated out of the 
 fundamental theory at higher
energies\footnote{
 The Standard
Model with the heavy top quark integrated out provides a phenomenologically
important realisation of this scenario. 
}, whose presence would cancel the apparent low-energy anomaly. 
Another example is the Green-Schwarz mechanism in string theory \cite{Coriano:2007fw, Coriano:2007xg}. 
 
Indeed, the presence of an anomaly in
the low-energy description can always be cancelled by a Wess-Zumino
term~\cite{PRESKILL1991323}, which is a higher-dimension operator in the
Lagrangian density of topological origin. Given that this is the case, one
might think that we should not impose anomaly cancellation as a condition,
since 
we are likely building an EFT only valid at low energies.
However, if one were to disregard the
constraint of anomaly cancellation, one should explicitly construct the
appropriate gauged Wess-Zumino terms to cancel all anomalies in the EFT, and
derive the phenomenological consequences of these terms (for 
example, they will generically entail new interactions of the SM gauge bosons\footnote{A
  textbook example of this occurs in the chiral Lagrangian describing pions,
  the physical degrees of freedom of QCD at low energies. There is a
  topological term in the action for this theory, which is the original
  Wess-Zumino-Witten (WZW) term~\cite{Wess:1971yu,Witten:1983tw}. Upon
  gauging electromagnetism, the WZW term contributes a dimension-5 operator in
  the Lagrangian proportional to $\pi^0 F \tilde{F}$, which facilitates the
  decay of the neutral pion $\pi^0$ to a pair of photons, thus playing a
  crucial r\^ole in the low-energy phenomenology of the theory. Generically,
  the addition of Wess-Zumino terms will involve similar operators coupling
  new scalar fields to the gauge bosons corresponding to the anomalous
  symmetries which are being matched by the Wess-Zumino terms, invariably
  changing the phenomenology of the gauge sector in such an EFT.}).  
  
Also, if anomaly cancellation in low-energy EFTs may be
ignored,  
it is at best curious that the SM cancels the anomalies of its gauge groups.
We strongly suspect that the SM is at most an EFT
description of fundamental physics, since it does not include dark matter,
have sufficient baryogenesis, or include gravity, for example.
And yet, the SM conspires to be an
anomaly-free, perfectly consistent renormalisable gauge field theory in and of
itself. Such a conspiracy might suggest that we should take
anomaly cancellation seriously when we try to go beyond the SM.

Furthermore, given an anomalous assignment of charges at low
energies, it is usually
difficult to know for certain that an appropriate set of beyond the SM chiral
fermions can indeed be written down and given suitably large masses in a
consistent framework\footnote{Even though a suite of Wess-Zumino terms can indeed
  always be written down in the low-energy EFT to cancel all
  anomalies, this does not guarantee that such operators can in fact arise
  (with the precise coefficients to cancel anomalies) in the low energy limit
  of a renormalisable quantum field theory defined in the UV\@. For instance, the
   fact that only certain Wess-Zumino terms are allowed is what 
  gives rise to monotonicity theorems along RG flows~\cite{Zamolodchikov:1986gt, Komargodski:2011vj}.}.
For many charge assignments, this will prove impossible.
It is pragmatic, therefore, to ensure anomaly cancellation without the
need for Wess-Zumino terms\footnote{Thanks
  to the topological nature of the 
  Wess-Zumino terms, their coefficients are typically not renormalised. 
  In this case, their 
  coefficients can be tuned to zero in the EFT in a radiatively stable 
  way.}, as this removes a potential obstacle to
finding an UV complete description of the EFT.

\paragraph{RH neutrinos:}
Supposing sufficient knowledge of the heavy fermions at high energies, then {\em specific}\/ violations of EFT anomaly cancellation
 are possible. The example of the SM$\nu_R$ shall
prove to be pertinent and pedagogical here: in the low-energy effective
theory below some high scale 
associated with the masses of RH neutrinos\footnote{The RH
  neutrino masses are often set to be around
$10^{11}-10^{13}$ GeV in 
order to explain the smallness of the neutrino masses 
(after the see-saw
mechanism has made the left-handed neutrinos very light).}, two of the ``SM
anomaly cancellation equations''
({\em i.e.\ the equations not including the RH neutrinos' charges}\/) will
 seem violated, but in a correlated 
manner. RH neutrinos
are a special case because, 
 being chiral fermions but SM singlets,
their mass terms are invariant under the SM symmetries.
It is hard to imagine how to give non-SM singlet chiral representations a large
mass in an UV anomaly-free theory without breaking electroweak symmetry
prematurely
(i.e.\ at a scale much above the empirically determined electroweak scale
around 100 GeV), since the Dirac mass term will necessarily require left-handed
particles and 
a vacuum expectation value of an electroweak non-singlet.  

In the following, we shall take anomaly cancellation as a useful guide for
beyond the SM model building. This surely motivates an exploration of
the space of 
solutions to the anomaly cancellation equations. We chart the space
of family-dependent anomaly-free charge assignments in the two cases: the SM and
the SM$\nu_R$.

In the following \S~\ref{sec:ACC}, we define conventions and write down the anomaly
cancellation conditions, noting pertinent properties of them that help organise our solutions. Then, in \S~\ref{sec:dio}, a Diophantine
analysis shows how the solutions to the anomaly cancellation equations may be efficiently 
indexed and written in a closed form for either one or two families of
non-zero $U(1)_F$ charges. For the case of three families, certain existence arguments are formulated using modular arithmetic.
Next, in \S~\ref{sec:com}, a computer program is described that efficiently
solves the anomaly cancellation conditions for all three families, including the more general case of the SM$\nu_R$.
Various checks upon its output are performed. Interesting properties of
the solutions are listed along with some examples. A particularly pertinent example case is then treated in detail in \S~\ref{yukawa}, namely the case in which the sets of $U(1)_F$ charges permit all Yukawa couplings at the renormalisable level.
We conclude in \S~\ref{sec:conc}.

\section{Anomaly Cancellation Conditions \label{sec:ACC}}

In this section we reproduce the system of anomaly cancellation conditions
(ACCs) which  we shall solve. We consider the SM$\nu_R$, of which the 
SM is a special case (all RH neutrino $U(1)_F$ charges set to zero). 
We shall also point out 
 some basic features of these equations which both our solution methods shall
 exploit. We begin by setting out our conventions.  

We write the SM fermionic fields as the following representations of $SU(3) \times
SU(2)_L \times U(1)_Y$: 
$$
Q\sim (3, 2, 1/6), \
L\sim (1, 2, -1/2), \
e \sim (1, 1, -1), \
u \sim (3, 1, 2/3), \
d \sim (3, 1, -1/3).
$$
In the SM$\nu_R$, we include RH
neutrino fields $\nu \sim (1, 1, 0)$. When discussing Yukawa couplings
later, we will consider the Higgs
doublet $H \sim (1, 2, -1/2)$.
Each fermionic field comes in three copies (families). We shall
discriminate between the different families' $U(1)_F$ charges by a family index
$i \in \{1,2,3\}$ where relevant. 
It will sometimes be convenient to refer to a generic fermionic irreducible 
representation of the SM gauge group (e.g.\ the left-handed quark doublet $Q$);
these we shall refer to as different ``species''.
Here, we consider extending the SM gauge symmetry to $SU(3) \times
SU(2)_L \times U(1)_Y \times U(1)_F$. Then the $U(1)_F$ charge of field $X$ under the
new $U(1)_F$ gauge symmetry is labelled 
by $F_X$, for $X \in \{ Q_i,L_i,e_i,u_i,d_i,\nu_i,H \}$ and $i \in
\{1,2,3\}$. While the SM gauge symmetries are flavour universal, this $U(1)_F$
symmetry will be allowed to have family-dependent couplings. 

There are six ACCs, arising from the six
(potentially non-vanishing) triangle diagrams involving at least one
$U(1)_F$ gauge boson. 
The $SU(3)^2 \times U(1)_F$ ACC is
\begin{equation}
\sum_{i=1}^3 (2F_{Q_i} - F_{u_i} - F_{d_i})=0, \label{su3squ1f}
\end{equation}
the $SU(2)_L^2 \times U(1)_F$ ACC is
\begin{equation}
  \sum_{i=1}^3 (3 F_{Q_i} + F_{L_i})=0, \label{su2squ1f}
\end{equation}
the $U(1)_Y^2 \times U(1)_F$ ACC is
\begin{equation}
  \sum_{i=1}^3 (F_{Q_i} + 3 F_{L_i} - 8 F_{u_i} - 2 F_{d_i} - 6
  F_{e_i})=0, \label{ysqu1f} 
\end{equation}
whereas the gauge-gravity ACC is
\begin{equation}
  \sum_{i=1}^3 (6 F_{Q_i} + 2 F_{L_i} - 3 F_{u_i} - 3 F_{d_i} - 
  F_{e_i} - F_{\nu_i})=0. \label{gravsqu1f} 
\end{equation}
In addition to these four linear equations, there are two ACCs which are
non-linear in the $U(1)_F$ charges, 
which correspond to triangle diagrams involving more than one $U(1)_F$ gauge boson. The 
$U(1)_Y \times U(1)^2_F$ ACC is the quadratic
\begin{equation}
  \sum_{i=1}^3 (F^2_{Q_i} - F^2_{L_i} - 2 F^2_{u_i} + F^2_{d_i} +
  F^2_{e_i})=0, \label{eqn:quad} 
\end{equation}
and finally the $U(1)^3_F$ ACC is the cubic
\begin{equation}
  \sum_{i=1}^3 (6 F^3_{Q_i} + 2 F^3_{L_i} - 3 F^3_{u_i} - 3 F^3_{d_i} - 
  F^3_{e_i} - F^3_{\nu_i})=0, \label{eqn:cubic} 
\end{equation}
where we have included the RH neutrinos. These six conditions
constrain eighteen $U(1)_F$
charges in the SM$\nu_R$: $F_X$, for each $X \in \{ Q_i,L_i,e_i,u_i,d_i,\nu_i\}$,
with $i \in \{1,2,3\}$.
The SM chiral fermion content is obtained by restricting to the special case $F_{\nu_i}=0\ \forall i \in \{1,2,3\}$ (thus there are fifteen $U(1)_F$ charges in the SM case). However, note that the SM ACCs are obtained by the less restrictive pair of conditions $\sum_i F_{\nu_i}=\sum_i F_{\nu_i}^3=0$, which can indeed be satisfied for non-zero RH neutrino charges.

We note that the RH
neutrinos do not enter into the ACCS except for the gauge-gravity and the
$U(1)_F^3$ ACCs  
(Eqs.~\ref{gravsqu1f},\ref{eqn:cubic}) because
they are Standard Model singlets. Thus, if one did not know of the existence
of the $U(1)_F$-charged RH neutrinos and one used the SM version of
the equations, one might be misled by these two ACCs. This should not be an
excuse for neglecting the ACCs while setting up one's theory however, since we
notice from Eqs.~\ref{gravsqu1f},\ref{eqn:cubic} that the violations of their
SM limit are specific and correlated. Furthermore, the four other ACCs
must still be satisfied for anomaly freedom in the UV.

Some important features of the ACCs and their solutions are:
\begin{enumerate}

\item {\bf Rational solutions:} we shall assume that the solutions to the ACCs
  are valued in the rationals, $\mathbb{Q}$. 
In a holographic setting, if the
boundary conformal field theory is \emph{finitely generated} (notationally, has a finite number
of fields in the path integral), then the bulk gauge group must be
compact\footnote{More precisely, any potentially non-compact groups must be
  contained within a larger unified gauge group that is compact;
much as how the electromagnetic gauge group
is not {\em necessarily}\/ quantised, but is embedded into the compact SM group,
$SU(3) \times SU(2)_L \times U(1)_Y$.}~\cite[Theorem 6.1]{Harlow:2018tng}. As
finite dimensional representations of a compact Lie group have charges on a
discrete weight lattice, we are then guaranteed to have rational charge
ratios.   
  Put another way, if the ratio of two charges is irrational, they will not
  fit into a unified, compact, semi-simple, non-abelian
  group.
  For instance, we may imagine that the $U(1)_Y \times U(1)_F$ part of the
  symmetry (which would otherwise suffer from Landau poles in the gauge
  coupling at some high energy scale) is in fact embedded in a unified
  gauge-symmetry, corresponding to a semi-simple gauge group $G$. 

\item {\bf Rescaling invariance:} since the ACCs,
  Eqs.~\ref{su3squ1f}-\ref{eqn:cubic}, are homogeneous polynomials in the
  eighteen 
  variables, one may rescale all charges that specify a solution by
  any rational number
\begin{equation}
 F_X \to c F_X,\ \forall X \in \{ Q_i,L_i,e_i,u_i,d_i,\nu_i\} , \;\;\;\; c
 \,\in\, \mathbb{Q} \label{rescale}
\end{equation}
and arrive at another solution. These rescaled solutions are not  
independent, because rescaling all charges is equivalent just to an overall rescaling of the gauge
coupling. Hence, solutions related by such a rescaling are in an equivalence
class. Moreover, this freedom to rescale means that rational solutions may be taken to be 
in the integers $\mathbb{Z}$ without loss of generality\footnote{We note
that irreducible representations of $U(1)$ are labelled by integers anyway because the group transformations are defined to be periodic with
period $2\pi$~\cite{woit}.}.
However, one may not be free to rescale charges by changing the gauge
coupling to {\em any}\/ degree:
there is growing evidence that gravity must be the weakest force in a
consistent theory of quantum gravity~\cite{ArkaniHamed:2006dz}. In practice,
this puts a bound on how low one can make any gauge coupling $g$ in units
of the charge. 
The Weak Gravity Conjecture is the claim that
the low-energy EFT will always have a cutoff of at least $g M_P$, and there
must be at least one charged particle with a mass below this. 
Applied to our
$U(1)_F$ gauge coupling $g_F$, if the field with the largest $U(1)_F$ mass-to-charge
ratio has mass $m$ and $U(1)_F$ charge $F_X$,
\begin{equation}
g_F F_X  >  \frac{m}{M_P} , \label{wgc}
\end{equation}
for example if the particle with largest mass-to-charge ratio is a top quark with
 mass $m$ on the order of $100$ GeV, 
$g_F F_X > \mathcal{O}(10^{-17})$. 
If the bound in Eq.~\ref{wgc} is not satisfied, then one must introduce additional heavy
degrees of freedom charged under the $U(1)_F$ group, or else risk the EFT breaking down
prematurely from quantum gravity corrections.  
We also note that there is an {\em upper}\/ bound on $g_F$ if we require
perturbativity. Assuming that there are no fields charged under $U(1)_F$ other than SM$\nu_R$
fermions\footnote{Vector fields charged under $U(1)_F$ would weaken the bound,
whereas $U(1)_F$-charged scalar fields would strengthen it.}, the $U(1)_F$ beta function may be phrased as
\begin{equation}
\frac{d \ln g_F}{d \ln \mu} = \frac{\sum_{X_i} \left( F_{X_i} g_F
  \right)^2}{24 \pi^2}, \label{beta}
\end{equation}
where $X_i$ are SM$\nu_R$ Weyl fermions, $F_{X_i}$ are their $U(1)_F$ charges and $\mu$
is the renormalisation scale in the 
minimal subtraction scheme. 
For perturbativity
we should have that\footnote{A significantly stronger bound may be obtained
  under the assumption that our model remains a good effective field theory
  all the way up until the Planck scale. In that case, demanding no Landau
  pole between the $Z^\prime$ scale and the Planck scale results in a bound
  that is a factor $1/\sqrt{\ln (M_P^2/M_{Z^\prime}^2)}$ stronger.} 
$ d \ln g_F / d \ln \mu <1 \Leftrightarrow |g_F| \sqrt{\sum_{X_i}
  F_{X_i}^2} < 2 \pi \sqrt{6}$.

\item {\bf Permutation invariance:} the ACCs are all invariant under
  permutations of 
  family indices within an individual species. Hence, we shall identify anomaly-free solutions up to
  permutations of families within each individual species (thus quotienting by
  the discrete group $S_3^{\otimes 5}$ for the SM case, which is of order $6^5=7776$). In practice this
  is implemented by choosing an ordering within each species. In what follows we choose:  
\begin{equation}
F_{X_1} \leq F_{X_2} \leq F_{X_3}\ \forall X \in \{ Q,L,e,u,d,\nu\}. \label{ordering}
\end{equation}
We note that this ordering choice means that $F_{X_1}$, $F_{X_2}$ and $F_{X_3}$ do
not necessarily correspond to the usual families defined by increasing mass of the
corresponding fermion within the species $X$. The usual ordering is then defined by
a permutation of $\{ F_{X_1},\ F_{X_2},\ F_{X_3} \}$, which will in general be
a different permutation for each $X$.
\end{enumerate}

The ACCs and their 
  solutions   are left   unchanged by the addition of fermions which are
  vector-like under the full SM$\times U(1)_F$ gauge group, since the
  left-handed and right-handed fermionic components cancel. Although this
  plays no r\^{o}le in our analysis, we note here that arbitrary numbers of
  such vector-like fermion representations may be added to our solutions and
  the resulting model will still be anomaly-free.

We note in passing that if one wants to solve simple $U(1)$ systems of ACCs with identical fermions,
where one 
allows the number of fermions to vary, the non-linear ACCs can be reduced to linear
equations, quickly yielding solutions~\cite{Batra:2005rh,davidTong}.
Here though, since we have
fixed the number of fermions (albeit with different numbers for two different
cases: the SM and 
SM$\nu_R$), and since these fermions transform in fixed (and different) representations of the other factors of the gauge group,
we must utilise different methods. In the following section, we demonstrate the use of methods
often employed to analyse such systems of Diophantine equations.

\section{Diophantine Analysis \label{sec:dio}}

In  this section we shall show that integer solutions to the system of ACCs (\ref{su3squ1f}-\ref{eqn:cubic}) can be efficiently indexed by specifying,
\begin{align}
&\text{For one family,} \;\;\;\;\; &&\{ F_Q  \,,\, F_{\nu} \}    \nn   \\
&\text{For two families,} \;\;\;\;\; &&\{ F_{Q+} \,,\,  F_{\nu+}    \} + \mathbbm{Z}^4   \nn  \\
&\text{For three families and }F_{\nu+}={\bar F}_\nu=0, \;\;\;\;\;&&\{ F_{Q+}  \, , \, \bar{F}_Q \,,\, \bar{F}_L \,,\, \bar{F}_e  \,,\, \bar{F}_d  \,,\, \bar{F}_u \}  \nn
\end{align}
where $\bar{F}_X = F_{X_3} + F_{X_2} - 2 F_{X_1}$ and $F_{X+}\equiv \sum_{i=1}^K
F_{X_i}$ for $K$ families. 

We begin by rewriting the linear ACCs Eqs.~\ref{su3squ1f}-\ref{gravsqu1f} in
terms of the  
  sum of $U(1)_F$ charges within a species: 
\begin{align}
 F_{u+} &= 4 F_{Q+} + F_{\nu+}  , \qquad &F_{d+}   &= - 2F_{Q+} - F_{\nu+},
                                                    \nonumber \\
F_{e+} &= - 6 F_{Q+} - F_{\nu+} , \qquad &F_{L+} &= - 3 F_{Q+}.
\label{eqn:lin}
\end{align}
For one family, we have $F_{X+}=F_X$ and
there is a unique solution for each $F_Q$ and $F_{\nu}$.
For two families, the sums $F_{X+} = F_{X_1} + F_{X_2}$ of each species are uniquely fixed as in Eq.~\ref{eqn:lin}, and there are infinitely many solutions for each difference $F_{X-}\equiv F_{X_1}-F_{X_2}$: but as these are in one-to-one correspondence with the set of four positive integers, they are easily enumerated to any desired $Q_{\rm max}$, as shown in Eqs.~\ref{eqn:twofamilySM} and~\ref{eqn:twofamilySMnuR}.

For three families, the sums $F_{X+}=F_{X_1} + F_{X_2} + F_{X_3}$ are fixed as in Eq.~\ref{eqn:lin}, and the nonlinear constraints reduce to a pair of quadratic Diophantine equations for $F_{X_{32}} = F_{X_3} - F_{X_2}$,
which are known to have finitely many solutions in the range of interest, $0 \leq F_{X_{32}} \leq \bar{F}_X$.

\subsection{One family (or several families with family-universal charges)}

If there is only one non-zero $U(1)_F$ charge per species, or several families where the
charges are all the same within a species\footnote{Or, indeed, only two
  families with non-zero (but identical within a species) charges.}, then we
have six integers 
$\{F_Q, F_u, F_d, F_e, F_L , F_{\nu} \}$ and four linear constraints. 
Once these linear constraints are imposed, the quadratic and cubic constraints are automatically satisfied. This can be understood physically from the anomalies---if there is only one family, then $U(1)_Y \times U(1)_F^2$ and $U(1)_F^3$ are not independent of the other anomalies.  
 
All solutions can be specified by two integers, say $F_Q$ and $F_{\nu}$, in terms of which the other charges are
\begin{align}
 F_u = 4 F_Q + F_\nu  , \;\;\;\; F_d   = - 2F_Q - F_\nu  ,  \;\;\;\; F_e = - 6 F_Q - F_\nu , \;\;\;\; F_L = - 3 F_Q. \label{one family}
\end{align}
Using $F_Q$ to index the solutions has the advantage that any $F_Q \in \mathbbm{Z}$ admits a solution. Had we instead specified, say, $F_L$, and solved the linear equations, we would have found that only $F_L \in 3 \mathbbm{Z}$ yields integer solutions. 

\paragraph{Examples:}

Note that if we set $F_{\nu}=0$ and decouple the RH neutrinos,
  the solution in Eq.~\ref{one family} reduces to gauging an additional
  hypercharge in a direct product such as in the Third Family Hypercharge
  model~\cite{Allanach:2018lvl}. Alternatively, if we set $F_{\nu}=-3F_Q$, the 
  solution in Eq.~\ref{one family} reduces to gauging $B-L$, baryon number
  minus lepton number within that family, as has appeared in
  Refs.~\cite{Alonso:2017uky, Bonilla:2017lsq}. 

\subsection{Two families} \label{sec:two families}

Moving on to the case of two non-trivial charges per species, we now have twelve integers $\{F_{Q_i}, F_{u_i} , F_{d_i}, F_{e_i}, F_{L_i} , F_{\nu_i} \}$, where $i=1,2$. 
As before, we can immediately apply the four linear constraints to remove four variables, although now the quadratic and cubic constraints are not automatically satisfied. 
However, there is still a simplification: the cubic equation reduces to a quadratic constraint---i.e.\ we find that the $U(1)_F^3$ anomaly is only independent if there are RH neutrinos in addition to the SM particles.  

\paragraph{Decoupling variables:}
By going to variables
\begin{equation}
 F_{X+} = F_{X_1} + F_{X_2}  , \;\;\;\; F_{X-} = F_{X_1} - F_{X_2}  ,
\end{equation}
we find that the linear conditions depend only on $F_{X+}$, and the nonlinear conditions depend only on $F_{X-}$. We can therefore fix all $F_{X+}$ in terms of $F_{Q+}$ and $F_{\nu+}$ as before, and then solve the remaining conditions:
\begin{align}
0 &=  F_{Q-}^2 + F_{d-}^2 + F_{e-}^{2} - F_{L-}^2 - 2 F_{u-}^2,   \label{two family 1} \\
0 &=  F_{\nu +} \left( 3 F_{d-}^2 + F_{e-}^2 - F_{\nu-}^2 - 3 F_{u-}^2   \right),  \label{two family 2}
\end{align}
which are now both quadratic.

\paragraph{Solving Diophantine equations:}
A {quadratic} Diophantine equation of the form
\begin{equation}
 x_1^2 + \sum_{k=2}^{N-1} n_k x_k^2 = x_{N}^2
\end{equation}
has an infinite number of solutions, which can be parameterised by
\begin{align}
x_j = \begin{cases}
 a_1^2 - \sum_{k=2}^{N-1} n_k a_k^2  ,  &j=1 \\
 2 a_1 a_j  ,  &2 \leq j \leq N-1  \\
 a_1^2 + \sum_{k=2}^{N-1} n_k a_k^2 ,  &j=N.
 \end{cases}
\end{align}
To see that this parameterisation provides a complete list of all solutions (up to rescalings), consider \emph{any} particular solution $\{ x'_j \}$. This solution will be generated by
\begin{align}
a_j = \begin{cases} 
 x_1' + x_N' , &j=1  \\
 x_j' ,  &2 \leq j \leq N-1,
 \end{cases}
\end{align}
up to a rescaling by $1/2(x_1+x_N)$, and so scanning over all $\{a_j\}$ will generate all possible solutions. 

In the present case, this allows us to parameterise the $F_{X-}$ when $F_{\nu+}=0$ in terms of four positive integers $\{ a, a_e, a_d, a_u \}$:
\begin{align}
F_{Q-} &= a^2 - a_d^2 - a_e^2 + 2 a_u^2  , \;\;\;\;  F_{L-} = a^2 + a_d^2 + a_e^2 - 2 a_u^2 ,  \nn \\
F_{d-} &= 2 a a_d, \;\; F_{e-} = 2 a a_e , \;\; F_{u-} = 2 a a_u ,  \label{eqn:twofamilySM}
\end{align}
and when $F_{\nu+}\neq 0$ in terms of four positive integers $\{ a, A, A_d, A_u \}$, where the parameterisation is now given by
\begin{align}
F_{Q-} &= a^2 - 4 A^2 A_d^2 - \left( A^2 - 3 A_d^2 + 3 A_u^2  \right)^2 + 8 A^2 A_u^2  ,  \nn \\
F_{L-} &= a^2 + 4 A^2 A_d^2 + \left( A^2 - 3 A_d^2 + 3 A_u^2  \right)^2 - 8 A^2 A_u^2  , \nn \\
 F_{\nu -} &= 2 a \left( A^2 + 3 A_d^2 - 3 A_u^2  \right) ,  \nn \\
 F_{e-} &= 2 a \left( A^2 - 3 A_d^2 + 3 A_u^2 \right), \nn \\
F_{d-} &= 4 a A A_d, \;\; F_{u-} = 4 a A A_u.  \label{eqn:twofamilySMnuR}
\end{align}
Scanning over these positive integers will generate a complete list of the $F_{X_-}$.

\paragraph{Example:}
One may obtain the well-known $L_\mu - L_\tau$ anomaly-free assignment of
charges~\cite{Heeck:2011wj,Altmannshofer:2014cfa,Altmannshofer:2015mqa} as a
particular solution within this general class of two-family solutions (where
we identify the first family fermions with the $U(1)_F$-uncharged family). If one
sets all of the quark charges to zero, then Eq.~\ref{one family} implies that
the remaining sums of charges all vanish, i.e.\
$F_{L+}=F_{e+}=F_{\nu+}=0$, and 
Eqs.~\ref{two family 1},~\ref{two family 2} reduce to a single non-trivial
equation, $F_{e-}^{2} = F_{L-}^2$, with $F_{\nu-}$ being unconstrained. Thus,
if we insist that the only non-zero charges are for two families of
leptons, 
we obtain solutions of the form
$(F_{L_2},F_{L_3},F_{e_2},F_{e_3},F_{\nu_2},F_{\nu_3})=(a,-a,a,-a,b,-b)$ for any two integers $a$ and $b$, from
which we recover the $L_\mu - L_\tau$ assignment either with ($b=a$) or
without ($b=0$) the inclusion of RH neutrinos.

\subsection{Three families}

Finally we consider the case of three non-trivial $U(1)_F$ charges per species, giving eighteen integers $\{F_{Q_i}, F_{u_i} , F_{d_i}, F_{e_i}, F_{L_i}, F_{\nu_i} \}$, where $i=1,2,3$. 
As before, we can apply the four linear constraints to remove four variables, and now the quadratic and cubic constraints Eq.~\ref{eqn:quad} and Eq.~\ref{eqn:cubic} are fully independent. 

\paragraph{Decoupling variables:}
With an analogous change of variables
\begin{equation}
 F_{X+} = F_{X_1} + F_{X_2}  + F_{X_3} , \;\;\;\; F_{X_{32}} = F_{X_3} - F_{X_2}  , \;\;\;\; \bar{F}_X = F_{X_3} + F_{X_2} - 2 F_{X_1} ,
\end{equation}
we find that the linear conditions depend only on $F_{X+}$, and the nonlinear conditions depend only on $F_{X_{32}}$ and $\bar{F}_X$. We can therefore fix all $F_{X+}$ in terms of $F_{Q+}$ and $F_{\nu+}$ as before, and then solve the remaining conditions:
\begin{align}
&3 \left( F_{Q_{32}}^2 + F_{e_{32}}^2 + F_{d_{32}}^2 - F_{L_{32}}^2 - 2 F_{u_{32}}^2   \right) + \left( \bar{F}_Q^2 + \bar{F}_e^2 + \bar{F}_d^2 - \bar{F}_L^2 - 2 \bar{F}_u^2 \right) = 0,   
\end{align}
and
\begin{align}
&9 \Big[ 6 \bar{F}_Q F_{Q_{32}}^2 + 2 \bar{F}_L F_{L_{32}}^2  + 3 ( 2 F_{\nu+} -  \bar{F}_d ) F_{d_{32}}^2 + ( 2 F_{\nu+} - \bar{F}_e ) F_{e_{32}}^2  \nn  \\
&\qquad - 3 (2 F_{\nu+} +  \bar{F}_u ) F_{u_{32}}^2 - (2 F_{\nu+} + \bar{F}_{\nu} ) F_{\nu_{32}}^2  \Big]   \nn \\
&= 6 \bar{F}_Q^3  + 2 \bar{F}_L^3 - 3 \bar{F}_d^3  - 3 \bar{F}_u^3 - \bar{F}_e^3   - \bar{F}_{\nu}^3 
- 6 F_{\nu+} \left[  3 \bar{F}_d^2 - 3 \bar{F}_u^2 + \bar{F}_e^2 - \bar{F}_{\nu}^2       \right].
\end{align}

\paragraph{Relabelling:}
Note that in the original variables, we had the freedom to relabel families. In these new variables, this is realised as the freedom to replace
\begin{equation}
 F_{X_{32}} \to - F_{X_{32}} , \;\;\;\; \bar{F}_X \to \bar{F}_X,
\end{equation}
or to replace
\begin{equation}
 F_{X_{32}} \to  \frac{F_{X_{32}} + \bar{F}_X}{2}    ,\;\;\;\; \bar{F}_X \to \frac{3 F_{X_{32}} - \bar{F}_X }{2}. 
\end{equation}
The former of these (together with the removal of cross terms from the
quadratic constraint) is the real motivation for our choice of new
variables. Crucially, this $\mathbbm{Z}_2$ parity of $F_{X_{32}}$ means that the
cubic equation can only depend on $F_{X_{32}}^2$ and \emph{not} $F_{X_{32}}^3$. We
need therefore only specify the six $\bar{F}_X$, and then we are left with a
pair of quadratic Diophantine equations for the $F_{X_{32}}$. These are more
difficult to solve than the previous two family case, because in general the
combination of $\bar{F}_X^2$ in the quadratic constraint and $\bar{F}_X^3$ in
the cubic constraint need not sum into an integer squared, so there need not
be a neat parameterisation.   

In the new variables, our ordering condition Eq.~\ref{ordering}
corresponds to
\begin{equation}
0 \leq F_{X_{32}} \leq \bar{F}_X
\end{equation}
for each species. In a finite range, a system of quadratic Diophantine equations has finitely many solutions, so at least each choice of the $\bar{F}_X$ labels a finite family of solutions, which can be found numerically. 


We can in fact say a little more than this. By applying basic modular arithmetic arguments to this pair of quadratics, we shall show that the sets of $\bar{F}_X$ charges which admit solutions for the $F_{X_{32}}$ can in fact be classified in the case where $F_{\nu+}=0$, and fall into two distinct classes. In the case of the SM$\nu_R$ with three families and no other constraints on the charges, we  find that the full solution space evades even a classification such as this, at least using our methods.

\paragraph{Existence of solutions:}
Consider parameterising the charges mod $3$.
One may deduce that
\begin{equation}
\bar{F}_X \equiv F_{X+} \quad (\text{mod~} 3), \label{variables mod 3}
\end{equation}
which follows the definitions of $\bar{F}_X$ and $F_{X+}$.\footnote{We thank Joseph Tooby-Smith for sharing with us this observation, and the resulting classification of Eq. \ref{three family classes}.} We now split our analysis into two cases, considering firstly the SM and then the SM$\nu_R$.

\subsubsection{SM \label{fnu0}}

In the case where $F_{\nu+}=0$, Eqs. \ref{eqn:lin} and \ref{variables mod 3} imply that
\begin{equation}
\bar{F}_L\equiv\bar{F}_e \equiv 0 \quad (\text{mod~} 3). \label{leptons mod 3}
\end{equation}
If we parametrise the remaining $\bar{F}$ variables modulo 3 by defining
\begin{equation}
\bar{F}_X = 3 n_X + r_X, \qquad n_X\in \mathbbm{Z} \text{~and~} r_X\in\{ -1, 0, 1\}, 
\end{equation}
then the quadratic ACC implies that
\begin{equation}
r_Q^2 + r_d^2 \equiv 2 r_u^2 \quad (\text{mod~} 3), \label{quad mod 3}
\end{equation}
and the cubic constraint turns out to be automatically satisified modulo $3$ (as can be seen by substituting in $r_X^3 = r_X$).
Eq. \ref{quad mod 3} then has the following solutions: either $r_Q=r_d=r_u=0$, which implies $\bar{F}_Q\equiv\bar{F}_d\equiv\bar{F}_u\equiv 0$ (mod $3$), or else each of $r_Q$, $r_d$, and $r_u$ are equal to $\pm 1$.

In fact, we can go further still and rule out some of these classes by now considering the cubic ACC modulo $9$. This implies the constraint
\begin{equation}
r_Q + r_d + r_u \equiv 0 \quad (\text{mod~} 3). \label{cub mod 3}
\end{equation}
This, together with Eq. \ref{quad mod 3}, admits only the solutions $r_Q=r_d=r_u=0$, $r_Q=r_d=r_u=+1$, and $r_Q=r_d=r_u=-1$. We can identify the latter two as corresponding to the same equivalence class of solutions, since it is always possible to perform a rescaling to set (say) $r_u=+1$.

Thus, solutions for $F_{X_{32}}$ only exist when\footnote{In fact, this proof holds not just in the SM case, but in the slightly more general case that we include three RH neutrinos with charges such that $F_{\nu+}=0$ and $\bar{F}_\nu\in 3\mathbbm{Z}$. Hence, we have included $\bar{F}_\nu$ in Eq. \ref{three family classes}. }
\begin{align}
  ( \bar{F}_u, \bar{F}_Q, \bar{F}_d, \bar{F}_e, \bar{F}_L, \bar{F}_\nu ) \;\; \in \;\;&  ( 3 \mathbbm{Z} , 3 \mathbbm{Z}, 3 \mathbbm{Z}, 3 \mathbbm{Z}, 3 \mathbbm{Z}, 3 \mathbbm{Z} ) , \nonumber \\
& ( 3 \mathbbm{Z}+1 , 3 \mathbbm{Z}+1, 3 \mathbbm{Z}+1, 3 \mathbbm{Z}, 3 \mathbbm{Z}, 3 \mathbbm{Z} ).  \label{three family classes}
\end{align}
In terms of efficiency, if we scan the six $\bar{F}_X$ from 1 to $3N$, this has
reduced the number of computations from $3^6 N^6 = 729 N^6$ to only $2 N^6$,
assuming $F_{\nu+}=0$ and $\bar{F}_\nu\in 3\mathbbm{Z}$.

\paragraph{Over-restrictions:}
Under certain conditions, there are \emph{no} solutions to the anomaly equations with only SM fermions. 
For instance, in Ref.~\cite{Ellis:2017nrp}, 
Ellis, Fairbairn, and Tunney
show that there are no SM solutions if:
\begin{itemize}

\item All RH quarks are uncharged,

\item At least one left-handed and one right-handed lepton is uncharged,

\item Two left-handed quark doublets have the same non-zero charge.

\end{itemize}
This is straightforward to see in our basis, as setting the RH quark charges to zero amounts to setting $F_{u,d+} = \bar{F}_{u,d} = 0 $, which then implies (by the linear constraints, Eq.~\ref{eqn:lin}) that all $F_{X+}$ are zero. Then, if we choose (without loss of generality) the zero lepton charges to be $F_{e_3} = F_{L_3} =0$, we have that $\bar{F}_{e} = F_{e+}$ and $\bar{F}_{L} = F_{L+}$ so these vanish as well. This leaves $\bar{F}_Q$ as the only non-zero $\bar{F}$, and consequently the cubic equation simplifies dramatically, to
\begin{equation}
\bar{F}_Q^3 = 9 \bar{F}_Q F_{Q_{32}}^2 .
\end{equation}
If two of the left-handed doublets, $F_{Q_i}$, then have the same charge, we can set $F_{Q_{32}} = 0$, and find that the only solution is $F_{Q_i} = 0$---so there can be no non-zero charge assignment as described in the third bullet point above.  
This is not the only set of conditions which leads to no possible SM solution, but it is a helpful example of how effectively the anomaly cancellation conditions can completely exclude \emph{all} charge assignments under certain conditions. 

\subsubsection{SM$\nu_R$}
Including the RH neutrinos, there are now more cases which admit solutions. 
We no longer have the simplification afforded by Eq. \ref{leptons mod 3}, with the quadratic ACC now being
\begin{equation}
r_Q^2 + r_d^2 + r_e^2  -  r_L^2  - 2 r_u^2  \equiv 0  \quad (\text{mod~} 3).
\end{equation}
Together with the cubic ACC (considered both modulo 3 and modulo 9), we obtain the set of constraints:
\begin{align}
r_Q^2 + r_d^2  + r_u^2  &\equiv  r_\nu ( r_\nu -  r_e )  &&(\text{mod~} 3),   \label{eqn:mod3quadratic} \\
r_L +  r_e + r_\nu  &\equiv 0  &&(\text{mod~} 3),     \label{eqn:mod3cubic} \\
r_Q -  r_L +   r_d  +  r_u &\equiv  r_{\nu+} ( r_e^2 - r_\nu^2 )  &&(\text{mod~} 3)     \label{eqn:mod9cubic}.    
\end{align}
In principle, one can proceed as above, and enumerate all solutions to Eqs.~\ref{eqn:mod3cubic},~\ref{eqn:mod3quadratic} and \ref{eqn:mod9cubic}.
However, for general $F_{\nu+} \neq 0$, we have not found efficiency savings such as those
found in \S~\ref{fnu0}. The case where $r_\nu = r_e$ is an exception (in which the right-hand-sides of Eqs.~\ref{eqn:mod3cubic},~\ref{eqn:mod3quadratic} and \ref{eqn:mod9cubic} all vanish); one can thence show that solutions can only exist in one of the following five classes,
\begin{align}
  ( \bar{F}_u, \bar{F}_Q, \bar{F}_d, \bar{F}_e, \bar{F}_L, \bar{F}_\nu ) \;\; \in \;\;&  ( 3 \mathbbm{Z} , 3 \mathbbm{Z}, 3 \mathbbm{Z}, 3 \mathbbm{Z}, 3 \mathbbm{Z}, 3 \mathbbm{Z} ) , \nonumber \\
& ( 3 \mathbbm{Z}+1 , 3 \mathbbm{Z}+1, 3 \mathbbm{Z}+1, 3 \mathbbm{Z}, 3 \mathbbm{Z}, 3 \mathbbm{Z} )  ,    \nonumber \\
& ( 3 \mathbbm{Z}+1 , 3 \mathbbm{Z}+1, 3 \mathbbm{Z}-1, 3 \mathbbm{Z}-1, 3 \mathbbm{Z}+1, 3 \mathbbm{Z}-1 ) , \nonumber \\
& ( 3 \mathbbm{Z}+1 , 3 \mathbbm{Z}-1, 3 \mathbbm{Z}+1, 3 \mathbbm{Z}-1, 3 \mathbbm{Z}+1, 3 \mathbbm{Z}-1 ) , \nonumber \\
& ( 3 \mathbbm{Z}+1 , 3 \mathbbm{Z}-1, 3 \mathbbm{Z}-1, 3 \mathbbm{Z}+1, 3 \mathbbm{Z}-1, 3 \mathbbm{Z}+1 ). 
\end{align}
Outside of the special case in which $r_\nu=r_e$
the space of solutions to the full three-family SM$\nu_R$ becomes harder to characterise.


In this generic three-family scenario including RH neutrinos, the problem
  ultimately reduces to a scan over integer solutions, albeit a scan only up
  to some maximum charges if we fix the values of the $\bar{F}_X$'s.
It is difficult to make any further progress solving the Diophantine
equations. Thus, in the generic situation, the development of 
an efficient 
computational search program is well-motivated. We describe such a program in
the next section.

\section{Computational Search \label{sec:com}}

In this section, we present a computational 
search over integers whose magnitudes are bounded by some user-defined $Q_\text{max}\in \mathbb{N}$.

\subsection{Efficient computation}
Blindly searching over all sets of
integers within this range and checking Eqs.~\ref{su3squ1f}-\ref{eqn:cubic} would be extremely inefficient: in the
SM$\nu_R$, we would
need to check six equations for $(2 Q_{max}+1)^{18}$ sets of $U(1)_F$ charges. 
If we take $U(1)_Y$ as an example, we can rescale the gauge coupling such that
the smallest hypercharge is one, in which case the maximum absolute
value of hypercharge is 6. Setting $Q_\text{max}$ to be the same value (6) would then require
checking the Eqs.~\ref{su3squ1f}-\ref{eqn:cubic} $1.0 \times 10^{20}$ times in order to find solutions
to the ACCs.
In order to make things more efficient, our computer program searches over automatically
ordered $U(1)_F$ charges and explicitly uses the four linear ACCs
Eqs.~\ref{su3squ1f}-\ref{gravsqu1f}, to reduce the number of sets to be searched
over by a factor of $7776(2 Q_\text{max}+1)^4$ for the SM, with an extra
reduction by a factor of 6 for the SM$\nu_R$. 
Further reductions result from scanning over only one representative from each equivalence
class of solution,
and from choosing the order of cycling through 
species
 in order to reduce
the number of operations. 

Sometimes in the cycling, the charge assignment of a
species $X$ exhibits $U(1)_F$ ``charge inversion symmetry'' (CIS)
where $\{F_{X_1}, F_{X_2}, F_{X_3}\} = \{-F_{X_1}, -F_{X_2}, -F_{X_3}\}$ taking
  into account the fact that the ordering does not matter. CIS charge
  assignments are of the form $\{
  -a,0,a \}$.
If, in the cycling, all species' $U(1)_F$ charges set so far are CIS (or indeed no
charges have yet been set), 
the next species' charges are chosen such that the number of positive charges
is less than or equal to the number of negative charges. This avoids 
cycling over both $F_X=\{-3, -2, -1\}$ and $F_X=\{1, 2, 3\}$ for instance,
which are in 
  the same equivalence class. Also, if the middle ordered charge is zero, then the
  magnitude of the third charge should be smaller or equal to the magnitude of the
  first. This avoids cycling over both $F_X=\{-1, 0, 2\}$ and $F_X=\{1,
  0, -2\}$, which are again in the same equivalence class. Once {\em all}\/ the
  $F_{X_i}$ 
  have been set, those assignments
  with a 
  greatest common divisor larger than 1 are identified by checking whether 
  all charges divide by the same prime number less than $Q_\text{max}$: if
  they do, they are removed
  from the list, since they are in the same equivalence class as an existing
  solution with smaller $U(1)_F$ charge magnitudes (which we take to be the
  representative of the equivalence class). 

Bearing these considerations in mind,
$F_{Q_1}$ is chosen first to cycle through the range 
$[-Q_\text{max},0]$.
Thus, $F_{Q_1}$ is chosen in these sets to be {\em
  negative semi-definite}. Solutions with positive $F_{Q_1}$ can be obtained from
these by multiplying all $U(1)_F$ charges in the solution by the same -1 factor because
of the 
rescaling invariance of the ACCs and they are thus in the same equivalence
class. Next, 
 $F_{Q_2}$ is chosen in the interval $[Q_1,0]$ (the upper bound is fixed by our
 requirement that the number of positive $U(1)_F$ charges should not be greater
 than the number of negative ones, as explained above).
Then $F_{Q_3} \in \left[Q_2,\ Q_\text{max}\right]$, checking
that $|F_{Q_3}|<|F_{Q_1}|$ if $F_{Q_2}=0$. 
Next, if the SM case is desired, all RH neutrino $U(1)_F$ charges are set to
zero. 
Otherwise, $F_{\nu_i}$ are cycled\footnote{The way in which the cycling is
  performed is much more detailed than our exposition. We refer interested
  readers to the source code of the computer program, which is available on \url{http://doi.org/10.5281/zenodo.3345889}~\cite{zenodo}.}.
$F_{e_1}$ and $F_{e_2}$ are cycled next, but
\begin{equation}
F_{e_3}=-6
F_{Q_+} - F_{\nu_+} - F_{e_1} -F_{e_2}
\end{equation} is fixed, as implied by Eq.~\ref{eqn:lin}. 
If
$|F_{e_3}|>Q_\text{max}$ or if $F_{e_3}<F_{e_2}$, the 
program reverts to the next inner-most cycling (i.e.\ $F_{e_2}$). 

The rest of the cycling proceeds in a similar manner to
that of $\{F_{e_1}, F_{e_2}, F_{e_3}\}$ 
(in the species order $u$, $L$, $d$) 
until the program tests the quadratic ACC
Eq.~\ref{eqn:quad}. 
 If
the quadratic ACC
is not satisfied, the inner-most
cycling is continued (i.e.\ $F_{d_2}$). When the quadratic is
satisfied, only then is the cubic ACC Eq.~\ref{eqn:cubic} tested. 
The design of the program thus reduces the amount of computation by
not calculating 
further when the $U(1)_F$ charges set so far are not consistent in some way;
either because the magnitude of a charge set is
necessarily larger than $Q_\text{max}$, or because the charges set are
inconsistent with the ACCs, or because they are in the same equivalence class
as some other set of charges that has already been tested (or will be tested). 

At the end of the process thus outlined, we are left with a list of all inequivalent
solutions with $U(1)_F$
charge magnitudes up to $Q_\text{max}$. 
Finally, successful sets of charges are output as well as other data such as
the number of ACC quadratics and cubics evaluated. 

\subsection{Results}

We now list some example results and their properties.
The full lists are
available in the form of labelled, easily read {\tt ASCII} files
for public use on 
{\tt Zenodo} at 
\url{http://doi.org/10.5281/zenodo.3345889}~\cite{zenodo} for $Q_\text{max}\leq
10$ in the SM and $Q_\text{max}\leq 10$ in the SM$\nu_R$. 
The program itself is also made available there if a larger value of $Q_\text{max}$ is desired by the user.

\begin{table}
\begin{center}
\begin{tabular}{|c|ccc|ccc|ccc|ccc|ccc|ccc|}\hline
 & $Q$ & $Q$ & $Q$ & $\nu$ & $\nu$ & $\nu$ & $e$ & $e$ & $e$ & $u$ & $u$ & $u$ & $L$ & $L$ & $L$ & $d$ &
                                                                              $d$
  & $d$ \\ \hline
1 &  0 & 0 & 0 & 0 & 0 & 0 & 0 & 0 & 0 & 0 & 0 & 0 & 0 & 0 & 0 & 0 & 0 & 0 \\
2 &   0 & 0 & 0 & 0 & 0 & 0 & 0 & 0 & 0 & 0 & 0 & 0 & -1 & 0 & 1&  -1 & 0 & 1 \\
3 &   0 & 0 & 0 & 0 & 0 & 0 & -1 & 0 & 1 & 0 & 0 & 0 & -1 & 0 & 1 & 0 & 0 & 0 \\
4 &   0 & 0 & 0 & 0 & 0 & 0 & -1 & 0 & 1 & -1 & 0 & 1 & 0 & 0 & 0 & -1 & 0 & 1 \\
5 &  -1 & 0 & 1 & 0 & 0 & 0 & 0 & 0 & 0 & 0 & 0 & 0 & -1 & 0 & 1 & 0 & 0 & 0 \\
6 &  -1 & 0 & 1 & 0 & 0 & 0 & 0 & 0 & 0 & -1 & 0 & 1 & 0 & 0 & 0 &  -1 & 0 & 1 \\
7 &  -1 & 0 & 1 & 0 & 0 & 0 & -1 & 0 & 1 & -1 & 0 & 1 & 0 & 0 & 0 & 0 & 0 & 0 \\
8 &  -1 & 0 & 1 & 0 & 0 & 0 & -1 & 0 & 1 & -1 & 0 & 1 & -1 & 0 & 1 & -1 & 0 & 1 \\
\hline
\end{tabular}
\caption{Inequivalent solutions to the anomaly equations for SM fermion
  content and $Q_\text{max}=1$. Each row shows an anomaly-free $U(1)_F$ charge
  assignment. Note that the charges
  within a species are labelled in increasing order from left to
  right and so the ordering does not reflect the family assignment.
  \label{tab:smMaxCh1}}
\end{center}
\end{table}

As an example, we display all eight solutions to the SM ACCs with $Q_\text{max}=1$ in
Table~\ref{tab:smMaxCh1}. Remembering that we have yet to identify each
$U(1)_F$ charge
with a particular family, we note that solution 3 of the table may correspond
to $L_{\mu} - L_{\tau}$, which has been the subject of some phenomenological
interest
recently~\cite{Heeck:2011wj,Altmannshofer:2014cfa,Altmannshofer:2015mqa,Banerjee:2018mnw}. All
of the solutions in the table are totally
CIS (i.e.\ every species is CIS).
For these CIS solutions, since $\sum_i F_{X_i}=0$ for
each species $X$, they automatically satisfy all four linear ACCs,
Eqs.~\ref{su3squ1f}-\ref{gravsqu1f}. Also, since $\sum_i F_{X_i}^3=0$, they automatically
satisfy the cubic Eq.~\ref{eqn:cubic}, and so the only non-trivial constraint on such a CIS charge assignment
is that 
it solves the quadratic ACC, Eq.~\ref{eqn:quad}. {\em A priori}, one may therefore suspect
that the majority of solutions will be CIS, since five out of the six ACCs are then solved ``for free'', but in
fact we find that such CIS solutions become much less frequent
as $Q_\text{max}$ is increased, at least until $Q_\text{max}=10$.

Even with $Q_\text{max}=1$, we already notice a new solution of interest for explaining the neutral current
 in $B$-decay data in the solution 5 of Table~\ref{tab:smMaxCh1}: i.e.\  the charge assignment (now listing the indices as actual  family indices in the weak eigenbasis)
$F_{Q_3}=1, F_{Q_2}=-1,
 F_{L_3}=1, F_{L_2}=-1$, with all other $U(1)_F$ charges vanishing. Once the $U(1)_F$ symmetry is spontaneously broken,
 provided there is some quark mixing between $b_L$ and $s_L$,
 this will result in a $Z^\prime$ boson coupling to $( \bar \mu \gamma_\mu P_L
 \mu)$, and 
 to $(\bar b \gamma^\mu
 P_L s)$. These couplings 
 are of the correct type~\cite{DAmico:2017mtc} to explain the neutral current
 $B$-meson 
 decay data,
 which disagrees at the $4\sigma$ level with SM predictions. It
 remains for future work to see whether the model has otherwise viable parameter space
 but if it does, this will constitute a very simple model (going only slightly
 beyond the simplified $Z^\prime$ models introduced in
 Refs.~\cite{Allanach:2017bta,Allanach:2018odd}) that explains the data and is
 free of anomalies. 

\begin{table}
\begin{center}
\begin{tabular}{|c|ccccc|} \hline
$Q_\text{max}$ & {\bf Solutions} & Symmetry & Quadratics & Cubics &Time/sec \\ \hline
1 & {\bf 8} & 8 & 32 & 8 & 0.0\\
2 & {\bf 22} & 14 & 1861  &161 & 0.0\\
3 & {\bf 82} & 32 & 23288  & 1061& 0.0\\
4 & {\bf 251} & 56 & 303949 & 7757& 0.0\\
5 & {\bf 626} & 114 & 1966248  & 35430& 0.0\\
6 & {\bf 1983} & 144 & 11470333 & 143171 & 0.2\\
7 & {\bf 3902} & 252 & 46471312 & 454767 & 0.6\\
8 & {\bf 7068} & 336 & 176496916 & 1311965& 2.2\\
9 & {\bf 14354} & 492 & 539687692 &3310802 & 6.7\\
10 & {\bf 23800} & 582 & 1580566538 & 7795283 & 20 \\
\hline \end{tabular}
\caption{Number of inequivalent solutions to the anomaly equations for SM fermion
  content and different maximum $U(1)_F$ charge $Q_\text{max}$. Each row contains the
  all-zero charge solution, as well as the solutions indicated in the rows above.
  The column marked
  ``Symmetry'' shows how  many of the solutions are CIS\@, which we can see soon becomes a minority as $Q_\text{max}$ gets larger.
  We also list the number of
  quadratic and cubic anomaly equations checked by the program, as well as the
  real time taken for computation on a {\tt DELL\texttrademark~XPS
    13-9350}
  laptop.  
  \label{tab:smNum}}
\end{center}
\end{table}
\begin{table}
\begin{center}
\begin{tabular}{|c|ccccc|} \hline
$Q_\text{max}$ & {\bf Solutions} & Symmetry & Quadratics & Cubics & Time/sec \\ \hline
1 & {\bf 38} & 16 & 144 & 38  & 0.0\\
2 & {\bf 358} & 48 & 31439 & 2829  & 0.0\\
3 & {\bf 4116} & 154 & 1571716 & 69421 & 0.1\\
4 & {\bf 24552} & 338 & 34761022 & 932736 & 0.6\\
5 & {\bf 111152} & 796 & 442549238 & 7993169 & 6.8\\
6 & {\bf 435305} & 1218 & 3813718154 & 49541883 & 56\\
7 & {\bf 1358388} & 2332 & 24616693253 &241368652  & 312\\
8 & {\bf 3612734} & 3514 & 127878976089 &978792750 & 1559\\
9 & {\bf 9587085} & 5648 & 558403872034& 3432486128 & 6584\\
10 & {\bf 21546920} & 7540 &  2117256832910&  10687426240 & 24748\\
\hline \end{tabular}
\caption{Number of inequivalent solutions to the anomaly equations for SM$\nu_R$
  fermion 
  content and different maximum $U(1)_F$ charges $Q_\text{max}$. Each row contains the
  all-zero charge solution, as well as the solutions indicated in the rows
  above.
The column marked
  ``Symmetry'' shows how  many of the solutions are CIS\@.
We also list the number of
  quadratic and cubic anomaly equations checked by the program, as well as the
  real time taken for computation in seconds on a modern {\tt
    DELL\texttrademark~XPS 13-9350} laptop.  
  \label{tab:smNuNum}}
\end{center}
\end{table}
For $Q_\text{max}>1$ in the SM, or even for $Q_\text{max}=1$ for the SM$\nu_R$,
the solutions are too numerous to list in this paper. We do, however, list the
number of 
solutions and some other properties in Tables~\ref{tab:smNum},~\ref{tab:smNuNum}. 
As previously advertised, we see that CIS solutions become
relatively less frequent in the list of solutions as $Q_\text{max}$
increases. Also listed are the number of times the program checked the
quadratic ACC and the number of times it checked the cubic ACC\@. We see that
the program runs quickly for low values of $Q_\text{max}$ on a modern
laptop. The time taken to run fits a $T(Q_\text{max})/\text{secs}= \exp\left(A+B Q_\text{max} + C
  {Q_\text{max}}^2\right)$ function well, with constants $A=-9.45$,
$B=1.38$, 
  $C=-1.40 \times 10^{-2}$ for the SM and $A=-12.3$,
$B=3.4$, 
  $C=-0.11$
for the SM$\nu_R$. 
For $Q_\text{max}$ much
higher than 10 in the SM, run-time may be an issue.
Higher
efficiency may be attained by only scanning for solutions in the two classes
identified analytically in Eq.~\eqref{three family classes}.  
For the particular pair of non-linear Diophantine equations that need to be
solved in these cases, the use of look-up tables contained within special {\tt
  Mathematica}\texttrademark\ functions may expedite the calculation. 
If $Q_\text{max}>10$ is desired in the SM$\nu_R$, for which we have not found analytic simplifications analogous to Eq.~\eqref{three family classes}, it may be
advantageous to adapt the program to run in parallel on many cores.
\begin{figure}
\includegraphics[width=0.5\textwidth]{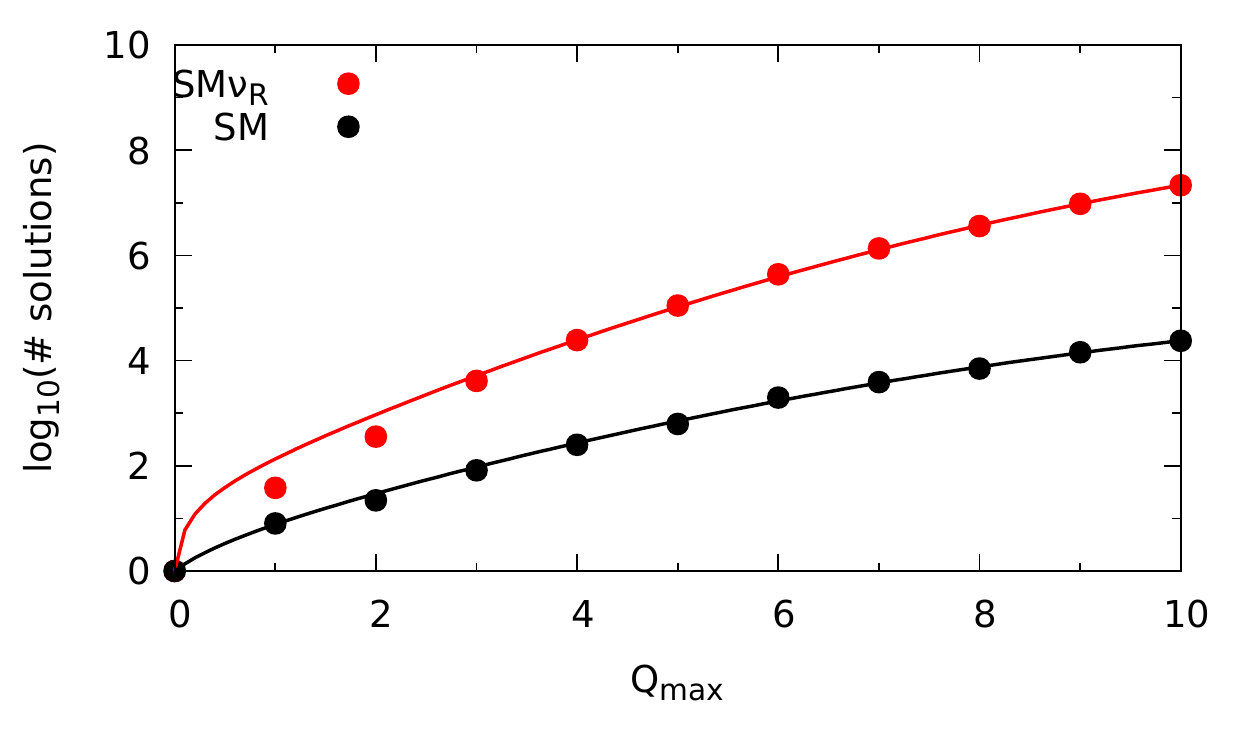}
\includegraphics[width=0.5\textwidth]{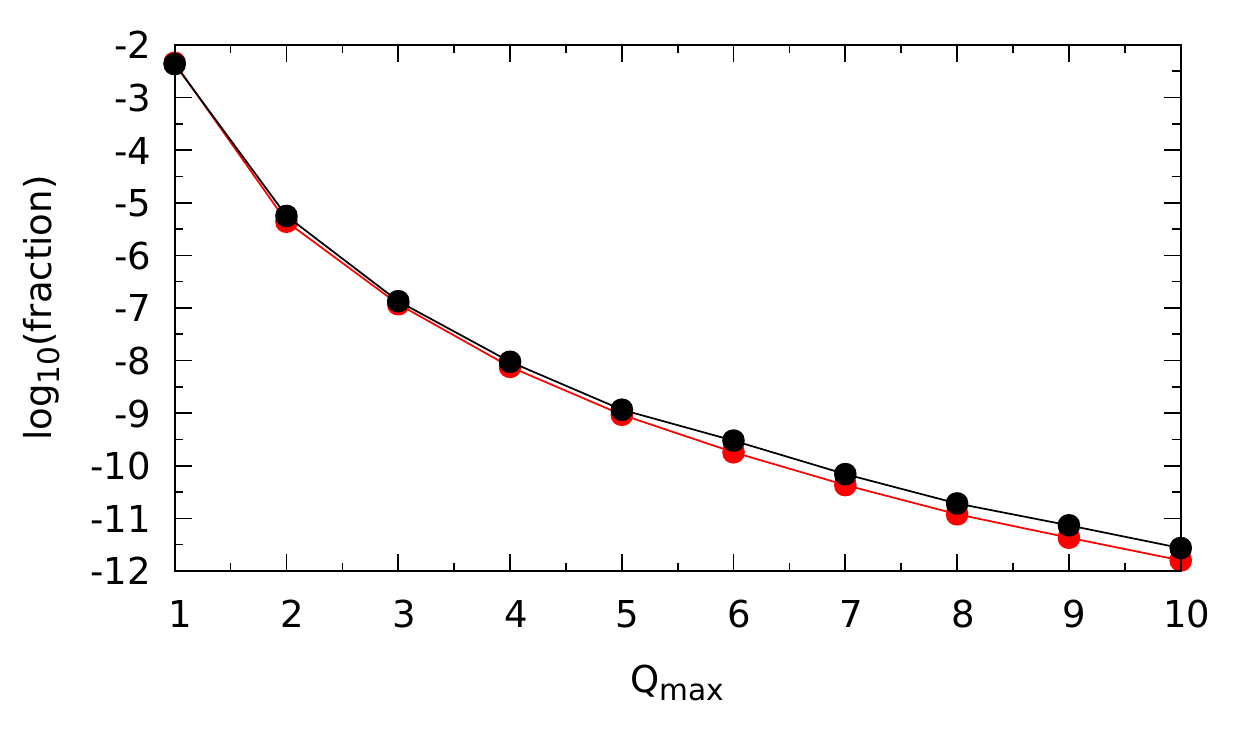}
\caption{\emph{Left}: The total number of inequivalent anomaly-free solutions with a given $Q_{\rm max}$, as
  tabulated in Tables~\ref{tab:smNum} and~\ref{tab:smNuNum}, together with the 
functions $1+a\exp(b Q_\text{max}+cQ_\text{max}^2)-a$ which are fit the growth of the
number of solutions: $a=22.5$, $b=2.0$, $c=-0.062$ for the SM$\nu_R$ and
$a=2.50$, $b=1.34$, $c=-0.043$ for the SM.
\emph{Right}: The fraction of all inequivalent charge assignments which is anomaly free for a given $Q_{\rm max}$, showing that imposing anomaly-freedom can lead to a drastic reduction in the available parameter space. 
\label{fig:sols}} 
\end{figure}
Fig.~\ref{fig:sols} shows the number of solutions as a function of
$Q_\text{max}$ graphically, along with some approximate numerical fits, for
the SM and the SM$\nu_R$. 

We note that, since the solutions for SM$\nu_R$ contain the solutions
where only one or two of the $\nu$ $U(1)_F$ charges are non-zero, these solutions also
correspond to the case of the SM plus only one or two RH neutrinos
(respectively).

\subsection{Queries of results}

\subsubsection{Testing against known solutions}
As a test of our program, 
we have checked our atlas of solutions for several anomaly-free $U(1)_F$ charge assignments that have been
previously 
identified and used in the literature: for example,
 $L_\mu-L_\tau$~\cite{Heeck:2011wj,Altmannshofer:2014cfa,Altmannshofer:2015mqa} is present in two forms: 
both in the SM
as in Table~\ref{tab:smMaxCh1} and in the SM$\nu_R$ with non-zero $\nu_R$ $U(1)_F$
charges.
\begin{table}
\begin{center}
\begin{tabular}{|c|ccc|ccc|ccc|ccc|ccc|ccc|} \hline
Model& $Q$ & $Q$ & $Q$ & $\nu$ & $\nu$ & $\nu$ & $e$ & $e$ & $e$ & $u$ & $u$ &
       $u$ & $L$ & $L$ & $L$ & $d$ & $d$ & $d$ \\ \hline
$L_\mu-L_\tau$ &  0  & 0 & 0 &  -1 &  0 &  1 &  -1 &0 &  1 & 0 &  0 & 0 &  -1 
                 & 0& 1 & 0 &  0  & 0 \\
TFHM &  -1 &  0 &  0 & 0 &  0 &  0 &  0 &  0 &  6 & -4 &  0 &  0 & 0 &  0  & 3 & 0 &0&  2 \\
$B_3-L_3$ & -1 &  0 &  0 & 0 &  0 &  3 &  0 &  0 &  3 & -1 &  0 &  0 &  0 &  0 &  3 & -1 &  0 &  0 \\
\hline\end{tabular}
\caption{$U(1)_F$ charges as output by the program for some example solutions
  present 
  in the literature. ``TFHM'' refers to the Third Family Hypercharge Model and
  $B_3-L_3$ to third family baryon minus lepton number. Note that the charges
  within a species are labelled in increasing order from left to
  right and so the ordering does not reflect the family assignment. Also note
  that the charges are multiplied by a rational constant in each case in order
  to get the traditional $U(1)_F$ charge assignments: -1/6 for the
  TFHM and -1/3 for $B_3-L_3$.
\label{tab:egs}} 
\end{center}
\end{table}
Third family baryon number minus (second or third family) lepton
number~\cite{Alonso:2017uky, Bonilla:2017lsq} 
is also present in the SM$\nu_R$, as is the 
Third Family Hypercharge Model~\cite{Allanach:2018lvl} in the SM\@. The
$U(1)_F$ charges
of these example charge assignments are shown in Table~\ref{tab:egs}. Several more solutions in the
literature were found in the output (all of the valid solutions sought for were
found), but here we omit them for brevity. 
On the other hand, as we discussed from an analytic perspective in \S~\ref{fnu0}, and as originally shown by Ellis, Fairbairn, and Tunney in Ref.~\cite{Ellis:2017nrp}, there are no SM solutions 
when (i) two of the $F_{Q_L}$ are non-zero and equal, (ii) there is at least one zero
$U(1)_F$ charge for both LH and RH leptons, and (iii) all RH quarks are
uncharged under $U(1)_F$. 
Searching our SM lists, we confirm that such solutions are absent,
providing another test of the program. We also confirm that when condition
(iii) is relaxed to ``all RH {\em down}\/ quarks are chargeless'', there are
still no 
non-trivial solutions found, agreeing with another result of
Ref.~\cite{Ellis:2017nrp}: that there are no rational solutions.

As a further test of our program ({and, indeed, as a cross-check on the results from our Diophantine analysis}), we check that 
Eq.~\ref{three family classes} applies for the subset of solutions with
$F_{\nu+}=0$. For the SM$\nu_R$ with
$Q_\text{max}=6$, out of the 435 305 solutions, 33 410 have $F_{\nu+}=0$ and, indeed,
we confirm that
all of these solutions fall into one of the two classes identified analytically in Eq.~\ref{three family classes} (once any solutions with 
$r_u=-1$ have been
rescaled such that $r_u$ is $+1$).

{While the full atlas of anomaly-free solutions which we list on {\tt
  Zenodo}~\cite{zenodo} might be intimidating for some readers, we point out
that imposing various phenomenology-motivated constraints on the possible
$U(1)_F$ charges is easy and fast. It will result in a cull of a large number of solutions
(e.g.\ 435305 for SM$\nu_R$ with $Q_\text{max}=6$), often down to a much
smaller list. We now demonstrate this further
through additional examples in \S~\ref{new solutions} and \S~\ref{yukawa}.}

\subsubsection{A few selected new solutions} \label{new solutions}
If we apply the less stringent
Ellis, Fairbairn, and Tunney conditions~\cite{Ellis:2017nrp} where, of the RH quarks, only
RH {\em down} quarks are fixed to be uncharged under $U(1)_F$ 
to the SM$\nu_R$ (they did not consider this case), we
find that  there are 20 solutions for $Q_\text{max}=6$, 
in all of which the RH
neutrinos are $U(1)_F$ charged. These solutions
therefore present a 
new example use case for our publicly available lists of solutions. 
They are reproduced in Table~\ref{tab:int}.
\begin{table}
\begin{center}
\begin{tabular}{|c|ccc|ccc|ccc|ccc|ccc|ccc|} \hline
  & $Q$ & $Q$ & $Q$ & $\nu$ & $\nu$ & $\nu$ & $e$ & $e$ & $e$ & $u$ & $u$ & $u$ & $L$ & $L$ & $L$ & $d$ &
                                                                              $d$
  & $d$ \\ \hline
1 &  -1 & 0 & 0 & 0 & 0 & 2 & 0 & 0 & 4 &-2 & 0 & 0 & 0 & 0 & 3 & 0 & 0 & 0 \\
2 &  -1 & 0 & 0 &-1 & 1 & 2 & 0 & 0 & 4 &-2 & 0 & 0 & 0 & 0 & 3 & 0 & 0 & 0 \\
3 &  -1 & 0 & 0 &-2 & 2 & 2 & 0 & 0 & 4 &-2 & 0 & 0 & 0 & 0 & 3 & 0 & 0 & 0 \\
4 &  -1 & 0 & 0 &-3 & 2 & 3 & 0 & 0 & 4 &-2 & 0 & 0 & 0 & 0 & 3 & 0 & 0 & 0 \\
5 &  -1 & 0 & 0 &-4 & 2 & 4 & 0 & 0 & 4 &-2 & 0 & 0 & 0 & 0 & 3 & 0 & 0 & 0 \\
6 & -1 & 0 & 0 &-5 & 2 & 5 & 0 & 0 & 4 &-2 & 0 & 0 & 0 & 0 & 3 & 0 & 0 & 0 \\
7 & -1 & 0 & 0 &-6 & 2 & 6 & 0 & 0 & 4 &-2 & 0 & 0 & 0 & 0 & 3 & 0 & 0 & 0 \\
8 &  -1 & 0 & 0 &-6 & 3 & 5 &-2 & 0 & 6 &-2 &-2 & 2 &-1 & 0 & 4 & 0 & 0 & 0 \\
9 &  -1 &-1 & 1 & 0 & 1 & 1 &-1 & 0 & 5 &-3 & 0 & 1 & 0 & 0 & 3 & 0 & 0 & 0 \\
10 &  -1 &-1 & 1 &-2 & 1 & 3 &-1 & 0 & 5 &-2 &-1 & 1 &-1 & 0 & 4 & 0 & 0 & 0 \\
11 &  -1 &-1 & 2 &-2 &-1 & 3 &-2 & 0 & 2 &-1 &-1 & 2 &-1 & 0 & 1 & 0 & 0 & 0 \\
12 & -1 &-1 & 2 &-3 & 1 & 2 &-6 & 0 & 6 &-2 &-1 & 3 &-5 & 0 & 5 & 0 & 0 & 0 \\
13 &  -1 &-1 & 2 &-3 &-1 & 4 & 0 & 0 & 0 &-1 & 0 & 1 &-1 & 0 & 1 & 0 & 0 & 0 \\
14 &  -1 &-1 & 2 &-3 &-1 & 4 &-1 & 0 & 1 & 0 & 0 & 0 &-2 & 0 & 2 & 0 & 0 & 0 \\
15 &  -1 &-1 & 2 &-3 &-1 & 4 &-3 & 0 & 3 &-2 & 0 & 2 &-2 & 0 & 2 & 0 & 0 & 0 \\
16 &  -1 &-1 & 2 &-3 &-1 & 4 &-4 & 0 & 4 &-3 & 0 & 3 &-1 & 0 & 1 & 0 & 0 & 0 \\
17 &  -1 &-1 & 2 &-3 &-2 & 5 &-6 & 0 & 6 &-3 & 1 & 2 &-5 & 0 & 5 & 0 & 0 & 0 \\
18 &  -1 &-1 & 2 &-5 &-1 & 6 &-6 & 0 & 6 &-3 & 1 & 2 &-5 & 0 & 5 & 0 & 0 & 0 \\
19 &  -2 &-2 & 3 &-4 & 0 & 6 &-2 & 0 & 6 &-3 &-1 & 2 &-2 & 0 & 5 & 0 & 0 & 0 \\
20& -2 &-2 & 3 &-4 & 0 & 6 &-2 & 0 & 6 &-4 & 0 & 2 &-1 & 0 & 4 & 0 & 0 & 0 \\
\hline\end{tabular}
\caption{$U(1)_F$ charges output by the program for solutions satisfying  Ellis,
  Fairbairn, and Tunney's less stringent conditions~\cite{Ellis:2017nrp} applied to the
  SM$\nu_R$ with $Q_\text{max}=6$. 
  Note that the charges
  within a species are labelled in increasing order from left to
  right and so the ordering does not reflect family assignment.\label{tab:int}} 
\end{center}
\end{table}
The phenomenology of the models in the table can be checked for desirable
properties: with suitable weak eigenbasis family assignments and assumptions
about fermion 
mixing (e.g.\ that $b_L$ and
$s_L$ mix when going to the mass eigenbasis and some other assumptions
involving lepton mixing), the first solution can be made to generate the
necessary couplings of a $Z^\prime$ to explain egregious neutral current
$B$-meson decay data, for instance. We note that only solution 14
satisfies the more stringent conditions where {\em all}\/ RH quarks are set to
be uncharged under $U(1)_F$ in a non-trivial way.

Some of the other solutions correspond to models which provide candidate solutions
to both the neutral current $B-$meson decay data and aspects of the fermion
mass problem. For 
example, consider the following
SM$\times U(1)_F$ solution, that appears in our atlas with $Q_\text{max}=4$:
\begin{equation}
\begin{array}{*{15}c}
Q_1  & Q_2 & Q_3 &  e_1 & e_2 & e_3 &  u_1 & u_2 & u_3 & L_1  & L_2 & L_3 & d_1 & d_2 & d_3 \\ 
0 & -3 & 3 & 0 & -3 & 3 & 0 & -1 & 1 & 0 & -4 & 4 & 0 & 0 & 0 \\
\end{array},
\end{equation}
where now the indices on the fields indicate family assignment in the weak
eigenbasis. Provided $b_L$ and $s_L$ mix, this model (once $U(1)_F$ is
spontaneously broken and the resulting $Z^\prime$ is integrated out) will
generate an $({\overline b_L} \gamma^\mu s_L) (\overline{\mu_L} \gamma_\mu
\mu_L)$ effective coupling of the kind indicated by fits to the neutral
 current $B-$meson 
decay data~\cite{DAmico:2017mtc}.
If we set the Higgs $U(1)_F$ charge equal to $+2$ in these units, then the only
renormalisable Yukawa coupling permitted by this pattern of charges is that of
the top quark. Presuming that the other Yukawa couplings arise as higher
dimensional operators after integrating out some UV physics (involving, say,
vector-like fermions), then the banning of all other Yukawa couplings at the
renormalisable level would naturally explain the fact that only the top Yukawa
coupling is of order one, as we will discuss more at the end of
  \S~\ref{yukawa}. This is yet another model of interest for further  
phenomenological study. In the following section, we discuss the implications of anomaly cancellation if we require 
that {\em all} of the electrically-charged fermion
Yukawa couplings are permitted at the renormalisable level.

\section{Constraints from a Renormalisable Yukawa Sector \label{yukawa}}

An especially well-motivated constraint on the $U(1)_F$ charges that one might like to impose,
which we have until now ignored, comes from the Yukawa sector. Naturally, the
vast majority of our anomaly-free solutions forbid the presence of SM-like
Yukawa interactions at the renormalisable level by $U(1)_F$ gauge invariance
(even if we exploit the freedom to give the Higgs a non-zero $U(1)_F$ charge
$F_H$, which does not spoil the ACCs because the Higgs is a scalar). So, a
natural question to ask is the following: which solutions in our anomaly-free
atlas permit all of the SM Yukawa couplings at the renormalisable level? In such
models, the fermions of the SM can acquire their masses in the same way as in
the SM\@. 

In this section, we will show that the constraints from a renormalisable
Yukawa sector turn out to be strong enough that we can identify the subspace
of such solutions completely analytically, using similar methods to
\S~\ref{sec:dio}, without the need to query the results of our computer
program. Nonetheless, even in this case, we find that our computer program is
a useful tool, because it efficiently organises the solutions by maximum
absolute charge. This ``simple ordering'' is difficult to arrive at using
the analytic parametrisation of the solution space.

\subsection{SM Yukawa interactions}

In the SM, one should generically allow all entries in each of the complex
three-by-three Yukawa matrices, $Y_e$, $Y_u$, and $Y_d$, including all of their
off-diagonal matrix elements (whose presence leads to the CKM and PMNS mixing
matrices). Requiring $U(1)_F$ gauge invariance then tells us that: 
\begin{enumerate}
\item The $U(1)_F$ charges for the SM fields $Q$, $u$, $d$, $L$, and $e$ must all be
  flavour universal in order for the off-diagonal terms to be $U(1)_F$
  invariant. Hence, the $U(1)_F$ charges for SM fields are fixed by the five variables $F_{X+}\equiv 3F_X$, with each $F_{X_{32}}$ and $\bar{F}_X$ being zero.
\item $F_{Q}-F_{u}=F_H$ for $U(1)_F$ invariance of the up-type quark Yukawa couplings,
\item $F_{Q}-F_{d}=-F_H$ for $U(1)_F$ invariance of the down-type quark Yukawa
  couplings,
\item $F_{L}-F_{e}=-F_H$ for $U(1)_F$ invariance of the charged lepton Yukawa couplings.
\end{enumerate}
In the case where $F_H=0$, this reduces to requiring $F_Q=F_u=F_d$, and $F_L=F_e$.

For all of the SM fermion fields, the $U(1)_F$ charges are fixed by Eq.~\ref{eqn:lin}, which implies 
\begin{equation}
F_{Q+}-F_{u+}=-(F_{Q+}-F_{d+})=-(F_{L+}-F_{e+})=-3F_{Q+}-F_{\nu+}.
\end{equation}
Hence, there are indeed anomaly-free solutions which permit all renormalisable Yukawa couplings, provided the Higgs has $U(1)_F$ charge
\begin{equation}
F_H=(-3F_{Q+}-F_{\nu+})/3,
\end{equation}
where recall $F_{Q+}=3F_Q$ in this scenario, and $F_{\nu+}=F_{\nu_1}+F_{\nu_2}+F_{\nu_3}$.
Hence, such solutions exist for any pair
of integers $(F_{Q+},F_{\nu+})$. 

\begin{table}
\begin{center}
\begin{tabular}{|c|c|ccc|ccc|ccc|ccc|ccc|ccc|} \hline
&& $Q$ & $Q$ & $Q$ & $\nu$ & $\nu$ & $\nu$ & $e$ & $e$ & $e$ & $u$ & $u$ & $u$ & $L$ & $L$ & $L$ & $d$ &
                                                                              $d$
  & $d$ \\ \hline
& 1&   0 & 0 & 0 & 0 & 0 & 0 & 0 & 0 & 0 & 0 & 0 & 0 & 0 & 0 & 0 & 0 & 0 & 0 \\
$F_H=0$& 2&  0 & 0 & 0 &-1 & 0 & 1 & 0 & 0 & 0 & 0 & 0 & 0 & 0 & 0 & 0 & 0 & 0 & 0 \\
& 3& -1 &-1 &-1 & 3 & 3 & 3 & 3 & 3 & 3 &-1 &-1 &-1 & 3 & 3 & 3 &-1 &-1 &-1 \\
\hline
& 4& -1 &-1 &-1 & 0 & 0 & 0 & 6 & 6 & 6 &-4 &-4 &-4 & 3 & 3 & 3 & 2 & 2 & 2 \\
& 5& -1 &-1 &-1 &-1 & 0 & 1 & 6 & 6 & 6 &-4 &-4 &-4 & 3 & 3 & 3 & 2 & 2 & 2 \\
$F_H \neq 0$& 6 & -1 &-1 &-1 &-2 & 0 & 2 & 6 & 6 & 6 &-4 &-4 &-4 & 3 & 3 & 3 & 2 & 2 & 2 \\
$F_{\nu+}=0$ & 7 & -1 &-1 &-1 &-3 & 0 & 3 & 6 & 6 & 6 &-4 &-4 &-4 & 3 & 3 & 3 & 2 & 2 & 2 \\
& 8 &  -1 &-1 &-1 &-4 & 0 & 4 & 6 & 6 & 6 &-4 &-4 &-4 & 3 & 3 & 3 & 2 & 2 & 2 \\
& 9 &  -1 &-1 &-1 &-5 & 0 & 5 & 6 & 6 & 6 &-4 &-4 &-4 & 3 & 3 & 3 & 2 & 2 & 2 \\
& 10 &  -1 &-1 &-1 &-6 & 0 & 6 & 6 & 6 & 6 &-4 &-4 &-4 & 3 & 3 & 3 & 2 & 2 & 2 \\
\hline
& 11 &   0 & 0 & 0 &-1 &-1 &-1 & 1 & 1 & 1 &-1 &-1 &-1 & 0 & 0 & 0 & 1 & 1 & 1 \\
& 12 &   0 & 0 & 0 &-4 &-4 & 5 & 1 & 1 & 1 &-1 &-1 &-1 & 0 & 0 & 0 & 1 & 1 & 1 \\
& 13 &  -1 &-1 &-1 & 1 & 1 & 1 & 5 & 5 & 5 &-3 &-3 &-3 & 3 & 3 & 3 & 1 & 1 & 1 \\
$F_H \neq 0$& 14 &  -1 &-1 &-1 & 2 & 2 & 2 & 4 & 4 & 4 &-2 &-2 &-2 & 3 & 3 & 3 & 0 & 0 & 0 \\
$F_{\nu+} \neq 0$ & 15 & -1 &-1 &-1 & 4 & 4 & 4 & 2 & 2 & 2 & 0 & 0 & 0 & 3 & 3 & 3 &-2 &-2 &-2 \\
& 16 &  -1 &-1 &-1 &-5 & 4 & 4 & 5 & 5 & 5 &-3 &-3 &-3 & 3 & 3 & 3 & 1 & 1 & 1 \\
& 17 &  -1 &-1 &-1 & 5 & 5 & 5 & 1 & 1 & 1 & 1 & 1 & 1 & 3 & 3 & 3 &-3 &-3 &-3 \\
& 18 &  -1 &-1 &-1 & 6 & 6 & 6 & 0 & 0 & 0 & 2 & 2 & 2 & 3 & 3 & 3 &-4 &-4 &-4 \\
\hline\end{tabular}
\caption{Inequivalent SM$\nu_R$ $U(1)_F$ charges as output by the program which allow
  all possible renormalisable Yukawa
  couplings for SM fermions, for $Q_\text{max}=6$. The first
  three solutions have 
  $F_H=0$ whereas the rest have $F_H \neq 0$. The fourth to the tenth solutions
have $F_{\nu+}=0$, {in which case the SM fermions must have $U(1)_F$ charge
  proportional to hypercharge}; the fourth solution {(in which the RH
  neutrinos have zero $U(1)_F$ charges and hence decouple from the ACCs)} is in
the same equivalence class as $U(1)_F$ charge equal to hypercharge for all
fermions.  
\label{tab:fil}} 
\end{center}
\end{table}
If we further wish that the SM Higgs doublet field be uncharged under $U(1)_F$
(for example, we 
may wish to avoid the contribution from the Higgs vacuum expectation value to
tree-level $Z-Z^\prime$ mixing that results otherwise), the sum of the
 RH neutrino $U(1)_F$ charges is fixed to be $F_{\nu+}=-3F_{Q+}$. 
In other words, with $F_H=0$,
 such solutions only exist (with the exception of the trivial $F_{X_i}=0$
 solution) in the SM$\nu_R$ with non-zero $U(1)_F$ charges for $\nu_R$, 
 not in the SM alone. 
In this simpler case, each lepton (including $\nu_R$) has $U(1)_F$ charge $-3F_Q$, where $F_Q$ is the charge of each quark.
 
 Conversely, if one seeks an anomaly-free $U(1)_F$ extension
 of the SM {\em without}\/ RH neutrinos (or, more precisely, an extension where $F_{\nu+}=0$), but with all renormalisable Yukawa
 couplings, then one is forced to give the Higgs a non-zero $U(1)_F$ charge, and the charges of the SM fermions must be proportional to their hypercharges. 

\subsection{Non-universal RH neutrino charges}

In any of these cases, the
$U(1)_F$ charges for $\nu_R$ don't necessarily also have to be
flavour-universal, since $\nu_R$ non-universality has no effect on the
electrically-charged lepton Yukawa couplings\footnote{We assume that neutrino mass generation
  requires further model building.}. If we allow non-universality in the RH
neutrinos, then the possible solutions allowing all SM Yukawa couplings are no longer
classified solely by the integer pair $(F_{Q+},F_{\nu+})$, but require in
addition two
more variables $\bar{F}_{\nu}$ and $F_{\nu_{32}}$, whose values are
constrained by the cubic ACC: 
\begin{equation}
9 ( \bar{F}_{\nu} + 2 F_{\nu+} ) F_{\nu_{32}}^2 =  \left(  \bar{F}_{\nu}  -  6 F_{\nu+} \right) \bar{F}_{\nu}^2 .
\label{eqn:YukawaCubic}
\end{equation}
Eq.~\ref{eqn:YukawaCubic} has rational solutions for $F_{\nu_{32}}$ if and only if the two brackets,
\begin{equation}
  \bar{F}_{\nu}  -  6 F_{\nu+}  =: A^2 , \;\;\;\;  \bar{F}_{\nu} + 2 F_{\nu+} =: B^2 ,
\end{equation}
have the property that $A/B$ is an integer. As any irrational factor of $A$ must be compensated by an identical factor in $B$, it follows that $A B$ is an integer also. Using our freedom to relabel families, we can take $A$ and $B$ to be the positive roots without loss of generality. 
Before giving a closed form expression for the solutions, let us comment on a few obvious branches of solutions\footnote{
As described above, we use our freedom to relabel the families to set $0 \leq F_{32} < \bar{F}$.
},
\begin{align}
&\bar{F}_{\nu} = F_{\nu_{32}} = 0 \;\;\;\; &&\implies \;\;\;\; &&F_1 = F_2 = F_3   \label{eqn:Yuk1} \\
&F_{\nu_{32}} = A = 0 \;\;\;\; &&\implies \;\;\;\; &&F_3 = F_2 = -4 F_1/5  \label{eqn:Yuk2}  \\
&\bar{F}_{\nu} = B = 0 \;\;\;\; &&\implies \;\;\;\; &&F_i = 0   \label{eqn:Yuk3} \\
&A = B, \;\; \bar{F}_\nu = 3 F_{\nu_{32}} \;\;\;\; &&\implies \;\;\;\; &&F_1 + F_3 = F_2 = 0  . \label{eqn:Yuk4}
\end{align}
Putting these aside, there are no further solutions in which either $F_{\nu_{32}}$ or $A$ or $B$ are zero. 
The cubic equation has one remaining branch\footnote{
Again, the positive root can be taken without loss of generality---the negative root corresponds to sending $F_{\nu_{32}}$ to $-F_{\nu_{32}}$.
},
\begin{equation}
A B = 2 F_{\nu_{32}} + \sqrt{ (2 F_{\nu_{32}})^2 -  3 (A^2 / 3 )^2  } 
\end{equation}
Demanding that the right hand side is an integer requires that $A^2$ is divisble by three\footnote{
This follows from solving $(2 F_{\nu_{32}})^2 - 3 (A^2 / 3 )^2 = Z^2$ in the manner described in \S~\ref{sec:two families}, which lets us write a complete list of solutions in the form: $2F_{\nu_{32}} = c_1^2 + 3 c_2^2$, $A^2/3 = 2  c_1 c_2$ and $Z = c_1^2 - 3 c_2^2$ , for every pair of integers $c_1$ and $c_2$. 
}.  
Every remaining solution can then be given in terms of two integers, $c_1$ and $c_2$,
\begin{equation}
F_{\nu +} = c_1^3 - 9 c_1 c_2^2 , \;\; \bar{F}_{\nu} = 6 ( c_1^3 + 3 c_1 c_2^2 ) , \;\; F_{\nu_{32}} = 6 c_2 ( c_1^2  + 3 c_2^2 )
\label{eqn:Yuk5}
\end{equation}
To prove that this generates all of the solutions, it suffices to show that any desired solution, $\{ F_{\nu+}', \bar{F}_{\nu}',  F_{\nu_{32}}'  \}$, can be written in this form. This is achieved by setting,
\begin{equation}
 c_1 = A' B' , \;\;\;\; c_2 =  A'^2 / 3 
\end{equation}
which reproduces the desired solution up to a rescaling of all neutrino charges by $8 A'^{3} B' $. This closed form therefore does not capture solutions in which $A$ or $B$ vanishes, which is why we separated those cases out explicitly. 
The set of Eqs.~\ref{eqn:Yuk1}-\ref{eqn:Yuk5} is the complete list of solutions to Eq.~\ref{eqn:YukawaCubic}. 

The disadvantage of this analytic solution is that it doesn't generate charge
 assignments in a way which is ordered simply, in terms of maximum absolute
 charge value. For instance, while
 $c_1 = c_2 = 1$ gives the simple assignment $F_1 = F_2 = -4$, $F_3 = 5$, the
 neighbouring $c_2= 2 c_1 = 2$ gives $F_1 = -113$, $F_2=-230$, $F_3 = 238$
 (up to rescaling). For this reason, it is still often more convenient to work
 with the results of the computer program, even when full analytic solutions
 are known.    

Our analytic results are borne out by the lists of solutions in our
atlas for solutions with $U(1)_F$ charge magnitudes up to $Q_\text{max}$.
Filtering the SM$\nu_R$ $Q_\text{max}=6$ solutions in our atlas with the
conditions 1-4 yields eighteen solutions, which are shown in Table~\ref{tab:fil}.
There are just three {equivalence classes of} solutions with $F_H=0$ (i.e.\
those avoiding tree-level $Z-Z^\prime$ mixing after
spontaneous $U(1)_F$ breaking). {The only one of these three with
  non-trivial charges for the SM fermions indeed has all quark charges equal
  to $F_Q$ and all lepton charges equal to $-3F_Q$. 
}
Of the other solutions, seven have the SM fermion
$U(1)_F$ charges being proportional to their hypercharges, {as follows from the $F_{\nu_R}$ charges being in
the pattern $\{ -a, 0, a \}$ (since then $F_{\nu+}=0$). The remaining
solutions are labelled by different values of $F_{\nu+}$ (relative to
$F_{Q+}$), with the pair $(F_{Q+},F_{\nu+})$ determining the other $U(1)_F$
charges in each case. Note that there may be multiple solutions given such a
pair, corresponding to different charge assignments for the RH
neutrinos which satisfy Eq.~\ref{eqn:YukawaCubic} (solutions 13 and 16 of Table~\ref{tab:fil}
are examples). 
}

It is worth emphasising that allowing all of the Yukawa couplings
to be present at the renormalisable level, as they are in the SM, is not
essential for beyond the SM model building. For example, it is reasonable (and for some purposes
desirable) to suppose that there is in fact no mixing in the
electrically-charged leptons, and that the PMNS mixing thus comes entirely
from the neutrinos. 
In such a set-up, in which the individual charged lepton numbers $U(1)_e$,
$U(1)_\mu$, and $U(1)_\tau$ would then be individually conserved, one no
longer has to require that the off-diagonal couplings in the charged lepton
Yukawa matrix $Y_e$ be $U(1)_F$ invariant. This means that one could relax the
flavour-universality constraint in the lepton fields $X_i\in \{L_i,e_i\}$ (but
not in the quark fields). Relaxing this assumption opens up many more
anomaly-free solutions in our atlas, including the $L_\mu - L_\tau$ solution
(in which all of the quarks are chargeless)~\cite{Heeck:2011wj,Altmannshofer:2014cfa,Altmannshofer:2015mqa}.

Another more generic 
possibility, which has been extensively explored,
is that not all fermions acquire their masses directly from renormalisable
Yukawa couplings. After all, while the top quark has an order-one Yukawa
coupling, the 
other fermions have much smaller couplings. Indeed, it is in many ways
attractive to explain the power-suppressed Yukawa couplings of all of the lighter SM
fermions by suggesting they arise from higher-dimensional operators in the SM
EFT, which can be achieved by {\em banning}\/ these couplings at the
renormalisable level. This idea dates back to Froggatt and Nielsen
\cite{Froggatt:1978nt}, and is an important part of many models that seek to
explain aspects of the flavour problem.



\section{Conclusions \label{sec:conc}}

We have analysed the six anomaly cancellation equations for the SM gauge group in a direct
product with a gauged $U(1)_F$ group, both with SM fermion content and with SM content plus (up
to) three
RH neutrinos. The fermionic $U(1)_F$ charges may depend
upon the family, a model building construct which is recently popular given
its potential to explain some interesting data in neutral current rare $B$
meson decays that is in tension with SM predictions. Many other uses of such
$U(1)_F$ gauge extensions have been employed in the literature. 
We have used Diophantine analysis to index the solutions, and
indeed these methods can produce the complete solution space in particular cases. 
It is clear from the analysis that there
is an infinite number of inequivalent 
(i.e.\ up to rescalings and permutations) integer
solutions to this set of equations.
In the case of the SM content with generic non-universal $U(1)_F$ charges, we find that the space of anomaly-free solutions is divided into two distinct classes which we have identified in Eq.~\ref{three family classes}.

To complement this Diophantine analysis, a computer program has been developed which scans over candidate solutions and
provides lists of successful ones up to some maximum absolute $U(1)_F$ charge
$Q_\text{max}$, in order to explicitly generate the solutions for the most general case.
The fact
that a computer program can be written to perform such a task is, perhaps, not
surprising. The surprising fact (at least to the authors) was the speed with
which such a program can be made to produce exhaustive lists considering the
fact that one is potentially scanning over 18 integers between $-Q_\text{max}$
and $Q_\text{max}$, where
$Q_\text{max}=10$. All runs took less than a day on a currently modern laptop, even for the
computationally most intensive run (e.g.\ 7 hours for SM$\nu_R$ with $Q_\text{max}=10$).  

To the best of our knowledge, an anomaly-free atlas such as we have provided has
not appeared 
in the literature before, although some handful of the individual solutions have
been found and examined. The solutions are legion (e.g.\ 435~305 for
 SM$\nu_R$ with $Q_\text{max}=6$) and so we find it likely that 
the majority of solutions
found have not 
appeared in the literature before.

In addition to its use as a look-up table which allows model builders
to check that their desired $U(1)$ charge assignments are anomaly-free, the
anomaly-free atlas can also inform the development of models in which only
some of the SM fermions have assigned charges, or in which only qualitative
features of the list is known. This is shown in our examples: one where we
require a 
renormalisable Yukawa sector and one where we demand the phenomenologically
motivated 
assignments of Ref.~\cite{Ellis:2017nrp}. The anomaly-free atlas provides an
efficient way to complete partial charge assignments in any gauged $U(1)$
extension of the SM or SM$\nu_R$.  

There are various useful extensions to the atlas that one can envisage.
One could chart
the anomaly-free solution space of other popular chiral fermion field contents
beyond SM$\nu_R$. 
For example, in the minimal supersymmetric standard model, fermionic partners of
the two Higgs doublets are included, and if these had non-zero $U(1)_F$ charges this
would change the anomaly cancellation equations and therefore change their solution space. Models with ``sterile
neutrinos'' may warrant the introduction of additional $\nu_R$ fields beyond
the three considered here, each with
associated $U(1)_F$ charges. One could also construct different anomaly-free
atlases for different symmetry breaking patterns, for example $SU(3)\times
SU(2)_L \times U_{F_1}(1) \times
U_{F_2}(1) \rightarrow SU(3)\times
SU(2)_L \times U(1)_Y$, where $F_1$ and $F_2$ are (generically) different
family charges for chiral fermions. 

The atlas of solutions is publicly
available as an aid and an inspiration to model builders and others, being
particularly easy to automatically scan through, looking for desirable
properties.  Various
solutions that have already been found in the literature are present, which
provides a positive validation check on the results. Another check comes from
the absence of two classes of rational $U(1)_F$ charge assignments in the SM 
which previous work has shown
to be anomalous~\cite{Ellis:2017nrp}. In the SM$\nu_R$ however, the  analysis
of Ref.~\cite{Ellis:2017nrp}
does not apply and we find new solutions within the same class. 
In general, there are a huge number of new
solutions, and already
at first glance several of them appear to be worthy of further
phenomenological study.

\section*{Acknowledgements}
We thank other members of the Cambridge Pheno Working Group, in particular J. Tooby-Smith, as well as D. Gvirtz, N.
Dorey, and D. Tong for their helpful advice and comments. 
SM is supported by an Emmanuel College Research Fellowship, and JD is supported by The Cambridge Trust.
This work has been partially supported by STFC consolidated grant
ST/P000681/1.

\bibliographystyle{JHEP-2}
\bibliography{articles_bib}

\end{document}